\documentclass[useAMS,usenatbib]{mn2e}

\usepackage{amsmath}
\usepackage{latexsym}
\usepackage{amssymb}
\usepackage{graphicx}

\usepackage{subfigure}

\newcommand{\detg}{{\sqrt{-g}}}

\newcommand{\dF}{{^{^*}\!\!F}}
\newcommand{\alf}{Alfv\'en}

\def\fps@figure{bp}%
\def\fps@table{bp}%
\def\fps@plate{bp}%
\def\eps@scaling{1.0}%
\newcommand\epsscale[1]{\gdef\eps@scaling{#1}}%
\newcommand\plotone[1]{%
 \centering
 \leavevmode
 \includegraphics[width={\eps@scaling\columnwidth}]{#1}%
}%
\newcommand\plottwo[2]{%
 \centering
 \leavevmode
 \columnwidth=.45\columnwidth
 \includegraphics[width={\eps@scaling\columnwidth}]{#1}%
 \hfil
 \includegraphics[width={\eps@scaling\columnwidth}]{#2}%
}%
\newcommand\plotfiddle[7]{%
 \centering
 \leavevmode
 \vbox\@to#2{\rule{\z@}{#2}}%
 \includegraphics[%
  scale=#4,
  angle=#3,
  origin=c
 ]{#1}%
}%

\newcommand\araa{\rmfamily{ARA\&A}}%
\newcommand\apj{\rmfamily{ApJ}}%
\newcommand\apjl{\rmfamily{ApJ}}%
\newcommand\aap{\rmfamily{A\&A}}%
\newcommand\mnras{\rmfamily{MNRAS}}%
\newcommand\prd{\rmfamily{Phys.~Rev.~D}}%
\let\la=\lesssim            
\let\ga=\gtrsim
\title[Disk-Jet Coupling in Black Hole Accretion Systems II:
  FFE Models]{Disk-Jet Coupling in Black Hole
  Accretion Systems II: Force-Free Electrodynamical Models}

\author[Jonathan C. McKinney and Ramesh Narayan]{Jonathan  C. McKinney$^{1}$
  \thanks{E-mail: jmckinney@cfa.harvard.edu (JCM); narayan@cfa.harvard.edu (RN)}
  and Ramesh Narayan$^{1}$\footnotemark[1]\\
  $^{1}$Institute for Theory and Computation, Center for
  Astrophysics, Harvard University, 60 Garden St., Cambridge, MA,
  02138
}

\begin{document}
  \date{Accepted 2006 October 19. Received 2006 September 17; in original form 2006 July 25}
  \pagerange{\pageref{firstpage}--\pageref{lastpage}} \pubyear{2006}
  \maketitle \label{firstpage}

  \begin{abstract}

    In paper I, we showed that time-dependent general relativistic
    magnetohydrodynamic (GRMHD) numerical models of accretion disks,
    although being highly turbulent, have surprisingly simple
    electromagnetic properties.  In particular, the toroidal current
    density in the disk takes the form $dI_\phi/dr \propto r^{-5/4}$.
    Guided by this simplicity, we use a time-dependent general
    relativistic force-free electrodynamics (GRFFE) code to study an
    idealized problem in which the accretion disk is replaced by an
    infinitely thin rotating equatorial current sheet.  We consider
    both an $r^{-5/4}$ current profile and an $r^{-1}$ profile, the
    latter corresponding to the paraboloidal model of Blandford \&
    Znajek (1977).  The force-free magnetosphere we obtain with the
    $r^{-5/4}$ current sheet matches remarkably well to the
    Poynting-dominated jet seen in GRMHD numerical models.  By
    comparing to the non-rotating force-free model studied in paper I,
    rotation is seen to lead to mild decollimation of the jet
    suggesting that hoop-stress forces nearly cancel centrifugal
    forces. In order to study the process that generates the corona
    and disk wind and destroys the ordered field in the corona in
    GRMHD numerical models, the force-free field with the $r^{-5/4}$
    current distribution is embedded in an accretion disk and followed
    in a GRMHD simulation.  Field at high latitudes is continuously
    transported to larger radii leaving a corona with only disordered
    field, while in the equator the turbulent field is accreted.
    Reconnection and magnetic stresses contribute to a magnetized,
    thermal wind without the aid of an ordered field threading the
    disk.
  \end{abstract}




  \section{Introduction}\label{introduction}

  Accretion disks around black holes have been recently studied using
  general relativistic magnetohydrodynamic (GRMHD) simulations that
  have been used to determine the structure of the disk, corona, wind,
  and Poynting-dominated jet \citep{dhk03,mg04}.  This paper continues
  a study of the coupling between these components of the accretion
  flow.

  The simulated accretion flows develop vigorous nonlinear turbulence
  driven by the magneto-rotational instability \citep{bh91,bh98}.  One
  might suspect that all fluid and electromagnetic quantities in the
  disk would therefore be chaotic.  However, as we showed in
  \citet{mn06}, the vertically integrated toroidal current exhibits a
  smooth and simple behavior \citep{mn06}.  Specifically, the toroidal
  current $I_{\phi}$ enclosed inside radius $r$ closely follows a
  power-law profile, \begin{equation}\label{iphi} \frac{dI_{\phi}}{dr}
  \propto \frac{1}{r^{2-\nu}} , \end{equation} with $\nu=3/4$.  This
  scaling is found to hold in a time-averaged sense but also at each
  instant of time.  It is also independent of the black hole spin or
  the initial conditions used for the simulation.  \citet{mn06} argued
  that the presence of the simple current distribution is the reason
  why GRMHD models have simple, nearly steady, collimated,
  magnetically-dominated polar regions which are qualitatively similar
  in structure to the force-free jet in the Blandford-Znajek (1977,
  BZ) model \citep{mg04,hk06,mckinney2006c}.

  The fact that the current distribution in GRMHD models is simple
  suggests that the accretion disk might, at some level of
  approximation, be idealised as a simple boundary condition in the
  equatorial plane.  Motivated by this possibility, we consider
  general relativistic force-free electrodynamical (GRFFE) models with
  equatorial boundary conditions derived from GRMHD numerical models
  of turbulent accretion disks.  Treating the accretion disk (or
  stellar surface) as a boundary condition is a commonly used device
  to obtain a basic understanding of jets and winds (see, e.g.
  \citealt{michel73,oka74,oka78,blandford76,bp82,lov86,hey89,nit91,li92,appl92,appl93,bes93,conone94,con94,con95a,con95b}).
  The GRFFE approximation of the magnetosphere of the accretion disk
  leads to simple tractable equations that can be solved analytically
  (or quasi-analytically) in certain regimes, such as for slowly
  rotating black holes \citep{bz77} or for special field geometries
  \citep{uzdensky2004,uzdensky2005}.

  The most famous GRFFE model that treats the disk as a boundary
  condition is the Blandford-Znajek (BZ) model \citep{bz77}, which
  involves a stationary disk with a fixed toroidal current density. BZ
  presented two solutions for slowly spinning black holes with an
  equatorial current sheet.  The first solution corresponds to a
  split-monopole field which is centered on the black hole; this
  solution is a general relativistic extension of the force-free pulsar
  magnetosphere solution of \citet{michel73} and has a power-law disk
  current of the form given in equation (\ref{iphi}) with $\nu=0$.
  The second solution corresponds to a paraboloidal field, as in
  \citet{blandford76}, and has $\nu=1$.

  While the monopole solution is not applicable to collimating jets,
  BZ's paraboloidal solution is a potentially interesting model of
  astrophysical jets.  However, the paraboloidal solution is
  incomplete because BZ do not ensure force balance across the
  boundary between field lines threading the black hole and those
  threading the equatorial accretion disk.  BZ identified this
  interface between the black hole and accretion disk as the {\it
  transition region}. Their estimate of the energy output of the
  paraboloidal solution may be qualitatively incorrect since a
  significant amount of the energy comes from the transition region
  where the disk contacts the ergosphere.  Since the work of
  \citet{bz77}, \citet{mac84,ga97} considered better-motivated
  force-free solutions in order to estimate the flux ratio of the hole
  and disk but did not account for the transition region between the
  black hole and disk.  \citet{mel06} recently developed some
  small-angle approximations for the polar jet for a Schwarzschild
  black hole, while \citet{uzdensky2005} studied closed field (no jet)
  geometries for a Kerr black hole.

  The difficulty of finding solutions to the GRFFE and GRMHD equations
  of motion has spawned a significant interest in developing
  time-dependent codes to solve these equations of motion.  Of
  relevance to this paper is that time-dependent force-free codes
  found that the monopolar BZ solution
  \citep{kom01,kom02a,kom02b,kom04} and the split-monopole BZ solution
  \citep{mckinney2006a} are stable. More recently force-free and MHD
  (in the highly magnetized limit) codes have been used to
  successfully study the pulsar magnetosphere in the regime that all
  field lines cross the light cylinder
  \citep{kom06,mckinney2006b,bucc06,spit06}.

  Our first objective in this paper is to use a time-dependent GRFFE
  code \citep{mckinney2006a} to evolve the black hole + equatorial
  current sheet system with disk rotation and an arbitrarily rapidly
  rotating black hole to self-consistently obtain the complete
  paraboloidal-type field solution.

  \citet{mn06} found that the current distribution and poloidal funnel
  field geometry in GRMHD simulations of accretion disks are not
  consistent with BZ's paraboloidal model with $\nu=1$ or BZ's
  monopolar model with $\nu=0$.  Instead, the best fit to the current
  distribution and the poloidal funnel field geometry is obtained with
  $\nu=3/4$.  Despite the apparent similarity between the $\nu=1$ and
  $\nu=3/4$ solutions, they are qualitatively different.  For example,
  the $\nu=1$ solution has a logarithmic divergence as a function of
  radius for the enclosed toroidal current.  On the other hand, the
  $\nu=3/4$ solution has a finite enclosed toroidal current.  Given
  the similarity between the GRMHD simulations and the $\nu=3/4$
  current distribution and resulting funnel field geometry, our second
  objective is to solve the time-dependent GRFFE equations of motion
  for this GRMHD-motivated model with $\nu=3/4$.

  In order to construct a force-free model that closely matches the
  GRMHD simulations, we require not only the current density in the
  equatorial sheet but also the angular frequency of the field lines
  at the equator.  The simple force-free models we described in
  \citet{mn06} had no rotation, but here we are interested in modeling
  the effects of rotation.  For this purpose we use the results of
  GRMHD simulations to construct a reasonable model of the field
  rotation frequency for the entire radial range from the horizon out
  well into the disk. \citet{mn06} found that the disk field angular
  frequency is Keplerian at large radii but makes a smoothly
  transition near the black hole to $\Omega_F/\Omega_H\sim 0.5$ for
  $a/M\gtrsim 0.4$ and $\Omega_F/\Omega_H\sim 1$ for $a/M\lesssim
  0.4$, where $\Omega_H$ is the angular frequency of the hole.  A
  GRFFE solution with this dependence for $\Omega_F$ and a current
  distribution of $\nu=3/4$ would be the closest a GRFFE model could
  come to the GRMHD simulations without including the vertical
  structure of the disk or the matter itself.  Also, with $\Omega_F$
  in the transition region modelled accurately, the energy output of
  the disk and the black hole can be measured and compared.  We study
  and compare the jet structure in the idealized GRFFE model with that
  found in GRMHD simulations of accretion disks \citep{mckinney2006c}
  and find remarkable agreement between the two.

  Our final objective is to demonstrate how the $\nu=3/4$ force-free
  solution changes as a result of the presence of matter and the
  magneto-rotational instability.  We choose the $\nu=3/4$ type
  stationary field geometry as the initial conditions and carry out a
  GRMHD simulation with an accretion disk superimposed onto the field.
  We find that the model develops a turbulent disordered disk, corona
  and wind, but the Poynting-dominated jet power output is unchanged
  relative to the GRFFE model.

  \subsection*{Paper Outline}

  In section~\ref{GRFFE}, we outline the numerical setup and the
  method used to obtain force-free solutions.  In
  section~\ref{parabz}, we study the $\nu=1$ paraboloidal force-free
  solution.  In section~\ref{fffourth}, we study the $\nu=3/4$
  force-free solution using GRMHD-motivated boundary conditions for
  the disk.  In section~\ref{GRFFE2GRMHD}, we discuss a GRMHD
  numerical model that is initialized with an accretion disk embedded
  with the $\nu=3/4$ force-free field.  We follow the destruction of
  the force-free field associated with the corona in GRMHD models.  In
  section~\ref{limitations}, we discuss the limitations of our
  calculations.  Finally, in section~\ref{conclusions}, we discuss the
  results and conclude.

  \subsection*{Units and Notation}

  The units in this paper have $G M = c = 1$, which sets the scale of
  length ($r_g\equiv GM/c^2$) and time ($t_g\equiv GM/c^3$).  The
  density scale is arbitrary, so values of the density are normalized by
  some fiducial field strength.  In order to convert to a physical value
  of the density for a given mass accretion rate, one must use the field
  strength as a function of black hole spin and mass accretion rate as
  given by GRMHD models such as described in
  \citet{mckinney2005a,mckinney2005b,mckinney2005c,mckinney2006c}.  The
  horizon is located at $r=r_+\equiv r_g(1+\sqrt{1-(a/M)^2})$).  For a
  black hole with angular momentum $J=a GM/c$, $a/M$ is the
  dimensionless Kerr parameter with $-1\le a/M \le 1$.

  The notation follows \citet{mtw73} and the signature of the metric
  is $-+++$.  Tensor components are given in a coordinate basis.  The
  components of the tensors of interest are given by $g_{\mu\nu}$ for
  the metric, $F^{\mu\nu}$ for the Faraday tensor, $\dF^{\mu\nu}$ for
  the dual of the Faraday, and $T^{\mu\nu}$ for the stress-energy
  tensor. The determinant of the metric is given by $\detg \equiv {\rm
  Det}(g_{\mu\nu})$. The field angular frequency is $\Omega_F\equiv
  F_{tr}/F_{r\phi}=F_{t\theta}/F_{\theta\phi}$.  The magnetic field
  can be written as $B^i=\dF^{it}$.  The poloidal magnetospheric
  structure is defined by the $\phi$-component of the vector potential
  ($A_\phi$). The current system is defined by the current density
  ($\mathbf{J}$) and the polar enclosed poloidal current
  ($B_\phi\equiv\dF_{\phi t}$).  The electromagnetic luminosity is
  $L\equiv -2\pi \int_\theta d\theta {T^{(EM)}}^r_t r^2\sin\theta$.
  See
  \citet{gmt03,mg04,mckinney2004,mckinney2005b,mckinney2005c,mckinney2006a}
  for details on this standard notation.  Kerr-Schild coordinates are
  used to avoid spurious reflections off the inner-radial boundary
  \citep{mg02}.

  \section{GRFFE Model with Disk Magnetosphere}\label{GRFFE}

  We study time-dependent GRFFE numerical models under the assumption
  that the turbulent accretion disk is well-modelled by an equatorial
  current sheet that is treated as a rotating conductor.  The GRFFE code
  described in \citet{mckinney2006a} is used to evolve the axisymmetric
  force-free equations of motion in the Kerr metric in Kerr-Schild
  coordinates.  This code has also been successfully used to study
  pulsar magnetospheres \citep{mckinney2006b} and the split-monopole
  problem \citep{mckinney2006a}.  The computational grid geometry is
  chosen to be the same as in \citet{mckinney2006c} with an inner radius
  inside the horizon and an outer radius at $10^3r_g$.  This grid
  focuses the resolution toward the polar axes at large radii in order
  to resolve the collimating jet.

  The resolution for all models is chosen to be $256\times 256$, and
  the solutions are well-converged compared to low resolution models
  except very close to the poles due to the coordinate singularity in
  spherical polar coordinates.

  \subsection{Obtaining a Stationary Solution in GRFFE}

  Steady state GRFFE numerical models with no discontinuities or
  surface currents above the disk surface are found by choosing
  boundary conditions determined by an analysis of the Grad-Shafranov
  equation (see, e.g., \citealt{bogo97,beskin97}). For solutions that
  pass through a light cylinder at some radius, one is required to fix
  the magnetic field component $B^\theta$ perpendicular to the disk
  and specify $2$ other constraints at the disk, viz., $E_\phi\equiv
  F_{t\phi}=0$, and $\Omega_F=\Omega_{\rm disk}$, the disk angular
  velocity. Note that one cannot leave both $B^\theta$ and $B^r$
  unconstrained, as this allows arbitrary redistribution of the
  currents in the disk.  Doing so gives too much freedom and the
  solution reverts to the split-monopole solution.  No boundary
  condition is applied on the black hole.

  For axisymmetric, stationary solutions, the frozen-in condition of
  ideal MHD implies that the field line velocity $v^i$ is completely
  determined by the field $B^i$ and field rotation frequency
  $\Omega_F$ (see equation 46 in \citealt{mckinney2006a}).  Thus,
  during the simulation the 3-velocity in the equatorial plane is set
  to agree with this condition.  Such a 3-velocity is generally
  time-like for points between the inner and outer light
  ``cylinders,'' but outside this region the 3-velocity can be
  space-like and so unphysical.  In the event that the 3-velocity is
  space-like, the Lorentz factor is constrained to a fixed large value
  as described in section 2.5 of \citet{mckinney2006a}.  This safety
  feature is necessary to handle the violent evolution seen early in
  the simulations, but it is not activated after the solution
  approaches a stationary state.

  In the work described here, the initial conditions of the
  simulations are chosen to correspond to a non-rotating, current-free
  (except at the equator) force-free solution with $B^\phi=0$.  For
  example, for the paraboloidal models, BZ's paraboloidal solution
  with $a/M=0$ is used to set the initial conditions.  For other
  power-law current profiles (e.g., $\nu=3/4$), \citet{mn06} describe
  how to calculate analytic solutions to the non-rotating problem.
  During the time evolution, $B^\theta$ at the disk is held fixed at
  its initial value, and the field angular velocity $\Omega_F$ is set
  to a physically motivated profile as suggested by the GRMHD models
  \citep{mn06}.  For models with rotating disks and/or rotating black
  holes, the initial non-rotating state is far from the steady state
  solution, so the model undergoes violent non-stationary evolution.
  Eventually, however, it relaxes to a steady-state solution.  All
  solutions thus found are necessarily stable to Eulerian axisymmetric
  perturbations.

  \subsection{Modelling the Transition Region}

  The transition region between the disk and the black hole must be
  treated carefully to avoid undesirable discontinuities in the
  magnetosphere away from the equatorial disk.  We now discuss how the
  vector potential ($A_\phi$, which is used to obtain $B^\theta$) and
  the field angular frequency ($\Omega_F$) are chosen.

  \subsubsection{Softened Vector Potential}

  If one uses an arbitrary vector potential to set the profile of
  $B^\theta(r)$ at the disk, then this boundary condition may not always
  allow for a stationary solution or one without discontinuities.  The
  problem is that we calculate $B^\theta$ from the vector potential
  ($A_\phi$) for an idealized problem with $a/M = 0$ or even $M = 0$,
  since these solutions are easy to obtain analytically.  The solutions
  typically consist of power-law profiles.  However, for general $M\neq
  0$ and $a/M \neq 0$, the radial eigenfunctions must change character
  near the black hole in such a way that the solution for the vector
  potential must become nearly constant close to the horizon.

  This property of the force-free equations near the horizon is related
  to the fact that solutions have a minimal energy if the black hole has
  a minimal tangential field on the horizon \citep{mt82a,mt82b}. For
  example, compared to the pure self-similar paraboloidal solution of
  \citet{blandford76}, BZ's paraboloidal solution for a disk around a
  black hole effectively softens the vector potential near the black
  hole to a monopolar form in order to account for the presence of the
  black hole.  The characteristic radius where the change occurs in the
  case of a non-rotating hole is $r\sim 2r_g$.  For rapidly rotating
  black holes the horizon becomes smaller but the ergospheric radius in
  the equatorial plane remains at $r=2r_g$.  So we expect generically
  that the softening of the potential will occur inside $r\sim 2r_g$ for
  all spins.

  In our work, we take the analytic vector potential solution
  corresponding to $M=0$, $a/M=0$ and soften it by replacing the
  radial coordinate $r$ as follows,
  \begin{equation}\label{transB}
    r\rightarrow \left( {r^2+r_{\rm transB}^2} \right)^{1/2} ,
  \end{equation}
  where $r_{\rm transB}$ is the transition radius.  This allows the
  solution to become monopolar near the horizon, thus modeling the
  effects of the mass and spin of the black hole.  Usually we set
  $r_{\rm transB}=2r_+$, but none of the results depend sensitively on
  this choice, as we have confirmed by trying various values in the
  range $r_{\rm transB}=(1.5 - 4) r_g$.

  \subsubsection{Modelling the Angular Velocity}

  The simplest force-free solutions have no discontinuities except as
  defined by the boundary conditions.  In time-dependent force-free
  simulations, discontinuities can appear near the boundary if no simple
  force-free solution exists with the chosen boundary condition.  This
  is something to be avoided.  Therefore, in order to guarantee a simple
  force-free solution, the boundary conditions must sometimes be
  modified to match smoothly onto the magnetosphere.  Even in
  quasi-analytic work it is useful to avoid discontinuities in the
  transition from the boundary conditions to the magnetosphere
  \citep{uzdensky2004,uzdensky2005}.

  In our models, the transition region where the disk meets the horizon
  requires careful treatment of the boundary condition in the disk. This
  is where the field line rotation switches from the profile set by the
  disk to the value required by the black hole space-time.  We have
  tried two different models of the transition region.  In one model, we
  take the field angular velocity profile to be given by
  \begin{equation}\label{model3}
    \Omega_F = \left\{\begin{matrix} \Omega_{\rm DBH}, & r<r_{\rm trans\Omega}, \\
    \Omega_K \left(1-\left(\frac{r_+}{r}\right)^3\right) + \Omega_{\rm
      DBH}\left(\frac{r_{\rm trans\Omega}}{r}\right)^{n_{\rm t}}, &
    r>r_{\rm trans\Omega}.\end{matrix}\right.
  \end{equation}
  The quantity $\Omega_{\rm DBH}$ denotes the value of $\Omega_F$ on
  the black hole horizon.  We choose $n_{\rm t}$ so that there is a
  smooth transition.  This model only applies for $a/M\ge 0$, and we
  find that values of $r_{\rm trans\Omega}$ in the range from $r_+$ to
  $2r_+$ give acceptable results.

  A second transition model is chosen to enforce $\Omega_F$ to be
  strictly constant inside the horizon and strictly fixed to the disk
  profile beyond some transition radius.  This model is given by
  \begin{equation}\label{model4}
    \Omega_F = \left\{ \begin{matrix}\Omega_{\rm DBH}, & r<r_0, \\
      \Omega_{\rm DBH} + A (r-r_0) + B (r-r_0)^3, & r_0\leq r \leq r_{\rm
    trans\Omega},\\ \Omega_{\rm disk}, & r>r_{\rm
    trans\Omega},\end{matrix}\right.
  \end{equation}
  where $\Omega_{\rm DBH}$ again denotes the value of $\Omega_F$ at
  the black hole horizon.  The values of $A$ and $B$ are chosen such
  that $\Omega_F$ and $d\Omega_F/dr$ are continuous at $r_0$ and
  $r_{\rm trans\Omega}$, where both radii are arbitrarily chosen.
  This model is suitable for any value of $a/M$.

  As in BZ77, we assume that on the horizon ($r=r_+$) the angular
  frequency of field lines that thread the disk match onto some {\it
    fraction} of the black hole angular frequency
  \begin{equation}
    \Omega_H\equiv \frac{a}{2r_+} .
  \end{equation}
  This matching of $\Omega_F$ between the disk and the magnetosphere
  of the black hole enforces the condition that no extra surface
  currents (i.e., jumps in the field) are present that would have been
  created by a jump in the field angular frequency at the interface
  between the disk and black hole magnetosphere.  To find force-free
  solutions with no discontinuities, we iteratively repeat the
  simulations with improved values of $\Omega_{\rm DBH}$ until the
  solution has a minimal discontinuity.

  \section{BZ Paraboloidal Solution}\label{parabz}

  In this section the GRFFE equations of motion are self-consistently
  evolved to find the stationary force-free magnetospheric solution
  corresponding to the toroidal current distribution of the paraboloidal
  solution of \citet{bz77}, viz.,
  \begin{equation}
    \frac{dI_{\phi}}{dr} = \frac{C}{r^{2-\nu}} ,
  \end{equation}
  with $\nu=1$.

  In the absence of rotation, the vector potential for the
  paraboloidal solution takes the form
  \begin{equation}
    \label{aphipara} A_\phi^{\rm (para)} =
    \begin{cases}
      +g(r,\theta) & \text{$\theta < \pi/2$} \\
      +g(r,\pi-\theta) & \text{$\theta > \pi/2$} ,
    \end{cases}
  \end{equation}
  \begin{equation}
    g(r,\theta)\equiv \frac{C}{2}[r f_- + 2M f_+(1-{\rm ln}f_+)] ,
  \end{equation}
  where $f_+=1+\cos\theta$, $f_-=1-\cos\theta$, and $M$ is the mass of
  the black hole.  As explained earlier, this non-rotating solution is
  used as the initial conditions for rotating models, i.e., we obtain
  the boundary condition on the field at the equatorial plane from the
  non-rotating solution, viz., $B^\theta=-A_\phi,r/\detg$, and we hold
  this fixed for all time.  We set $C=1$ throughout the paper, and
  this normalizes all energy densities and field strengths.

  The BZ paraboloidal solution is based on the non-gravitational
  solution found initially by \citet{blandford76}.  BZ hypothesized
  that the solution for the poloidal field is independent of the
  velocity profile of the disk (see eq. 3.10 in that paper).
  \citet{blandford76} noted in section~3iii that gravity will add
  corrections but, in fact, even without gravity there are corrections
  for any significant amount of disk rotation.  Keplerian disks in
  particular have substantially different solutions from the
  non-rotating case since the Keplerian speed is a non-negligible
  fraction of the speed of light near the black hole.  In the
  discussion that follows, we show that the \citet{blandford76} and BZ
  solution for $\Omega_F$ and the energy output only applies when the
  disk is slowly rotating.  We then calculate the correct solution for
  $\Omega_F$ and the power output for a Keplerian disk for the case of
  both slowly and rapidly rotating black holes.

  We model the transition region using equation (\ref{model4}) with
  $r_0=r_+$ and $r_{\rm trans\Omega}=2r_+$.  We have also tried $r_{\rm
    trans\Omega}=1.2 r_+$ and we find only small quantitative differences,
  though there are additional surface currents in the equatorial regions
  near the horizon because of the sharper transition in the $\Omega_F$
  profile.

  \subsection{Slowly Rotating BH and Non-Rotating Disk}

  \begin{figure}
    \includegraphics[width=3.3in,clip]{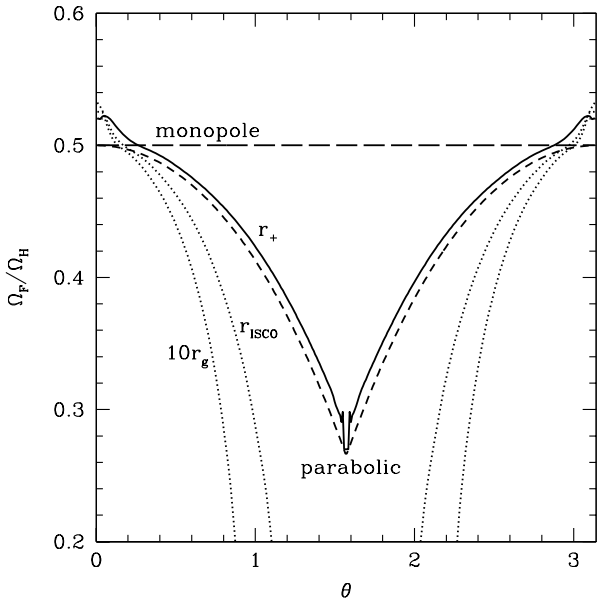}

    \caption{Field angular frequency per unit black hole
      angular frequency ($\Omega_F/\Omega_H$) vs. $\theta$ in radians on the
      horizon ($r=r_g$, solid line) and at two other radii ($r=r_{\rm
      ISCO}\approx 5.997 r_g$ and $r=10r_g$, dotted lines).  The
      results correspond to a GRFFE simulation of the paraboloidal
      solution ($\nu=1$) at time $t=1200t_g$ for a non-rotating disk
      and a slowly spinning black hole with $a/M=0.001$.  Also shown
      are the Blandford-Znajek monopole solution (long-dashed line)
      and paraboloidal solution (short-dashed line) on the horizon.
      Note the excellent agreement between the GRFFE numerical
      solution and BZ's analytical model. } \label{omegapara001}
      \end{figure}

  The first numerical model we describe has $a/M=0.001$, i.e., a very
  slowly rotating black hole, and we assume a non-rotating disk,
  $\Omega_{\rm disk}=0$.  We choose $\Omega_{\rm DBH} = 0.27\Omega_H$
  as in the BZ solution.  Figure~\ref{omegapara001} shows
  $\Omega_F/\Omega_H$ for the converged steady state solution at time
  $t=1200t_g$ at three different radii: the horizon at $r=r_+$, the
  inner-most stable circular orbit (ISCO) at $r=r_{\rm ISCO}$, and
  $r=10r_g$.  For comparison, the monopole and paraboloidal solutions
  on the horizon are also shown.

  The profile of $\Omega_F/\Omega_H$ from the numerical solution is very
  close to the analytical BZ solution.  The small offset is due to our
  model of the transition region, where one would expect to obtain BZ's
  solution only if allowing a discontinuity at the horizon as in their
  solution.  There is a small residual jump near the equator in
  $\Omega_F$ due to not choosing the most optimal value for $\Omega_F$
  on the horizon. Also, close to the poles $\Omega_F/\Omega_H$ deviates
  from the expected value of $\Omega_F/\Omega_H\approx 1/2$ due to the
  coordinate singularity, though this feature has a negligible impact on
  the solution at large radii due to the collimation of field lines
  toward the polar axis.

\begin{figure}
    \includegraphics[width=3.3in,clip]{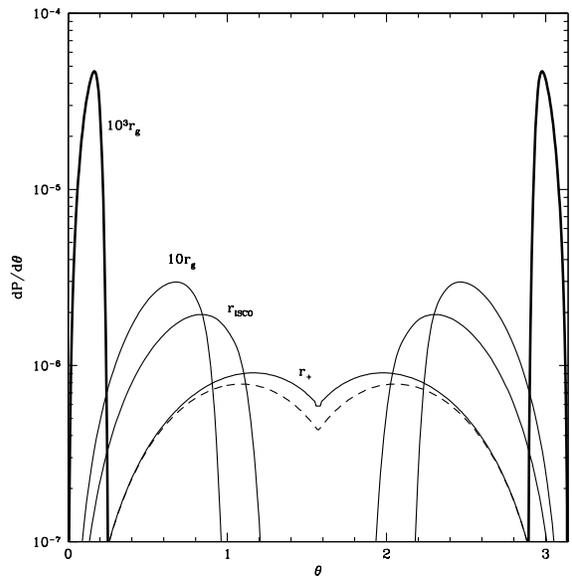}

    \caption{Angular density of the electromagnetic power output
      ($dP/d\theta$) vs. $\theta$ in radians at four radii, $r=\{r_+,
      r_{\rm ISCO}, 10r_g, 10^3r_g\}$.  The results are for the same
      GRFFE simulation shown in Figure~\ref{omegapara001}. The dashed line shows the
      analytic solution for the power output at the horizon according
      to the BZ paraboloidal model.  Note the excellent agreement
      between the numerical GRFFE result and BZ's analytical model. }
      \label{energyfluxpara001} \end{figure}

  Figure~\ref{energyfluxpara001} shows the angular density of the
  electromagnetic power output
  \begin{equation}
    \left\langle\frac{dP}{d\theta}\right\rangle = 2\pi r^2
    \left\langle{-T^{(EM)}}^r_t\right\rangle ,
  \end{equation}
  where $-{T^{(EM)}}^r_t$ is written in a coordinate basis in
  Boyer-Lindquist or Kerr-Schild coordinates.  The total integrated
  electromagnetic power at radius $r$ is obtained by integrating over
  angle,
  \begin{equation}
    P = \int_0^\pi \frac{dP}{d\theta} \sin\theta d\theta .
  \end{equation}

  The integrated electromagnetic power output at each of the four radii,
  $r=\{r_+, r_{\rm ISCO}, 10r_g, 10^3r_g\}$, is $P = \{1.4, 1.6, 1.6 ,
  1.6 \} \times 10^{-6}$, respectively, in units in which the radial
  field on the horizon at the poles is unity; that is, we scale the
  power as $(dP/d\theta)/B_0^2$ and $P/B_0^2$, where $B_0$ is the radial
  field strength on the horizon at the polar axis at the end of the
  simulation.  Normalized in the same way, BZ's paraboloidal solution
  gives a power of $P\approx 1.2\times 10^{-6}$ on the horizon
  \citep{bz77}.  Clearly, our numerical solution is in good
  agreement. The small differences arise because our numerical solution
  for $\Omega_F$ is slightly larger and the power on the horizon scales
  as $P\propto \Omega_F(\Omega_F-\Omega_H)$.  We find that along field
  lines the value of $\Omega_F$ is constant to within $5\%$, as required
  for an axisymmetric, stationary solution.

  Figure~\ref{energyfluxpara001} shows that the numerical profile for
  the angular power density agrees well with the analytical BZ
  solution. The slight difference is due to the introduction of the
  transition region between the disk and black hole.  Note that the
  power becomes progressively more collimated with increasing radius.
  The peak in the power density lies at only $\theta\approx 10^\circ$
  by a radius of $r=10^3 r_g$.

  One can also compute the average efficiency ($\bar{\epsilon}$) over
  the horizon, following equation 7.6 in \citet{bz77}:
  \begin{equation}\label{avgeff}
    \bar{\epsilon} = \frac{\int_0^\pi \sin\theta d\theta
      \left(\frac{dP}{d\theta}\right) }{\int_0^\pi \sin\theta d\theta
      \left(\frac{dP}{d\theta}\right)\left(\frac{\Omega_H}{\Omega_F}\right) }.
  \end{equation}
  Effectively, this is an estimate of $\Omega_F/\Omega_H$, weighted by
  the power density.  This quantity has a maximum value of $50\%$ for
  the monopolar solution.  Our numerical solution gives an efficiency of
  $39\%$, which is similar to the value of $38\%$ for the analytical BZ
  paraboloidal solution.

  In summary, the GRFFE numerical solution for the force-free
  magnetosphere surrounding a slowly-spinning black hole and a
  non-rotating disk agrees well with the analytical solution of BZ.

  \subsection{Slowly Rotating Black Hole and Keplerian Disk}

  We now describe a second numerical GRFFE model, which is identical
  to the previous model except now the disk rotates at the Keplerian
  frequency, $\Omega_{\rm disk} = \Omega_K = 1/(r^{3/2}+a)$ in
  Boyer-Lindquist coordinates.  Because the Keplerian speed near the
  black hole is much larger than the black hole rotation speed, it
  leads to significant changes in the field around the horizon.  The
  transition from the Keplerian disk to the relatively slow black hole
  speed forces the electromagnetic energy to be accreted rather than
  extracted from the black hole.  The integrated electromagnetic power
  output at the radii $r=\{r_+, r_{\rm ISCO}, 10r_g, 10^3r_g\}$ is $P
  = \{-0.001, 0.36, 1.42 , 2.04 \}$, respectively, in the same units
  as before.  For such slowly spinning black holes with rapidly
  rotating disks, the power output on the horizon depends sensitively
  on how we model the transition region between the disk and black
  hole.  In particular, the negative power on the horizon results from
  a complicated set of currents in the magnetosphere that provide a
  black hole-disk connection.  The substantial power at large radii
  primarily reflects the power output of the rotating disk and there
  is very little power from the black hole itself.  For all attempted
  models of the transition region, we find additional surface currents
  in the magnetosphere above the disk.

  \subsection{Rapidly Rotating BH and Keplerian Disk}

  \begin{figure}
    \includegraphics[width=3.3in,clip]{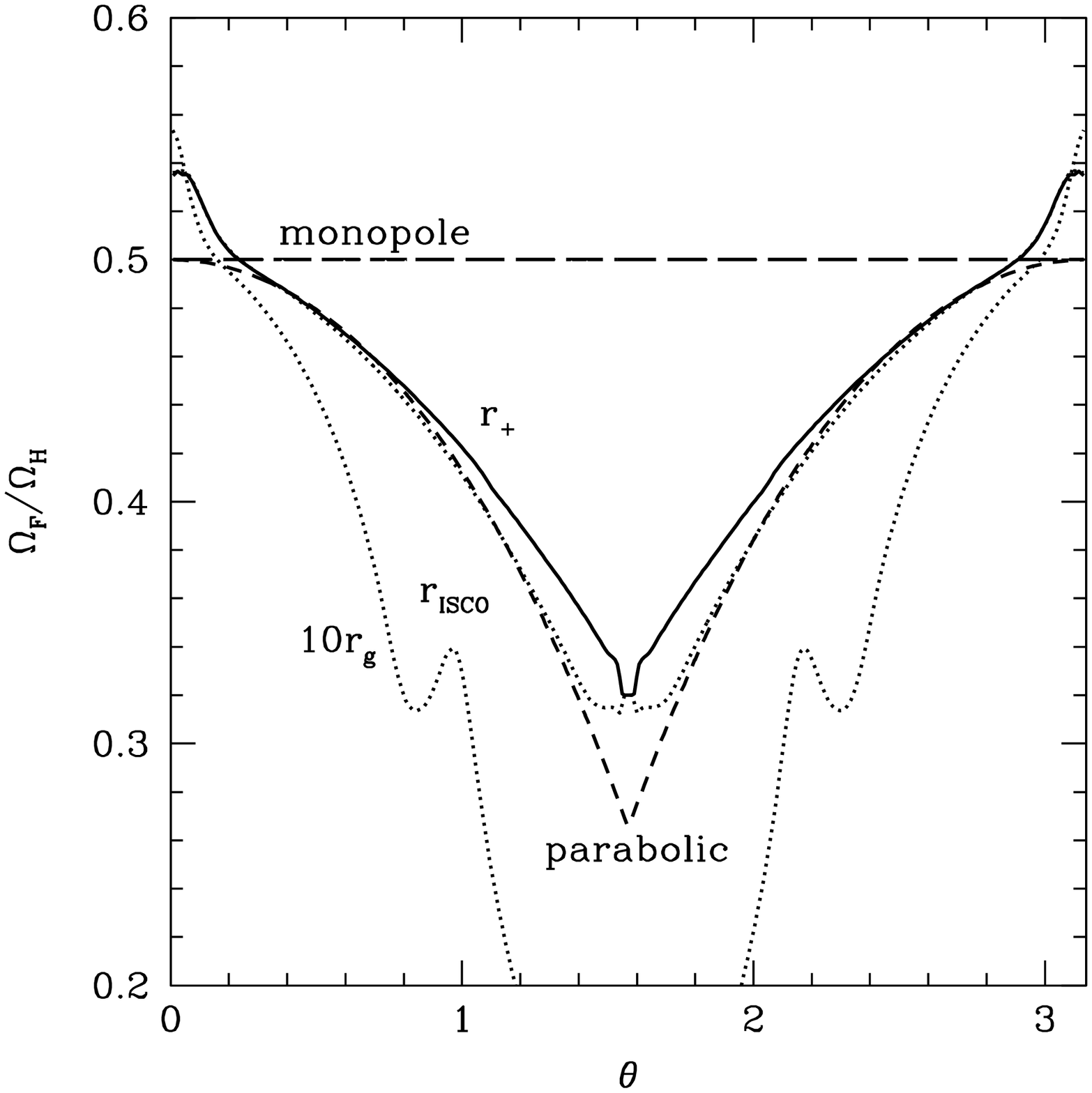}

    \caption{Field angular frequency per unit
      black hole angular frequency ($\Omega_F/\Omega_H$)
      vs. $\theta$ in radians on the horizon ($r=r_g$, solid line) and
      at two other radii ($r=r_{\rm ISCO}\approx 2.321 r_g$ and
      $r=10r_g$, dotted lines).  The results correspond to the
      paraboloidal solution ($\nu=1$) at time $t=1200t_g$ for a
      Keplerian disk around a rapidly spinning black hole with
      $a/M=0.9$.  Also shown are the Blandford-Znajek monopole
      solution (long-dashed line) and paraboloidal solution
      (short-dashed line) on the horizon.  The agreement between the
      numerical GRFFE model and the analytical BZ model is not as good
      as for a slowly rotating black hole (Figure~\ref{omegapara001}), but the deviations
      are still fairly mild. } \label{omegapara9} \end{figure}

  Finally, we consider a rapidly rotating black hole with $a/M=0.9$,
  surrounded by a Keplerian disk ($\Omega_{\rm disk}=\Omega_{\rm K}$).
  If we directly use BZ's paraboloidal solution for a slowly rotating BH
  to setup the disk boundary condition for the currents, only for
  $a/M\ll 1$ were we able to find a solution without discontinuities
  near the horizon.  In order to account for the additional effects of
  rapid spin, we had to soften the vector potential by replacing the
  radial coordinate as described in equation (\ref{transB}).  This was
  not required for slowly spinning models.

  With this model of the transition region we were able to find a
  solution with $\Omega_{\rm DBH} = 0.32\Omega_H$, which is a slightly
  larger value than for BZ's paraboloidal solution for a slowly spinning
  black hole.  Figure~\ref{omegapara9} shows $\Omega_F/\Omega_H$ at
  $t=1200t_g$ at three different radii and also shows the monopole and
  paraboloidal solutions on the horizon for comparison.

  \begin{figure}
    \includegraphics[width=3.3in,clip]{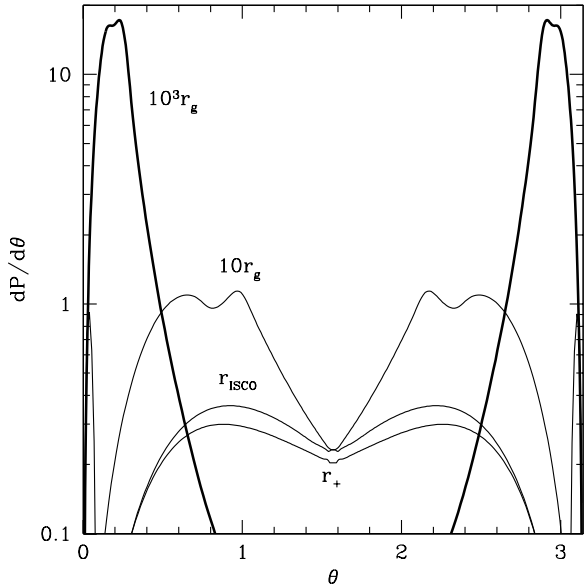}

    \caption{Angular density of the electromagnetic power output
      ($dP/d\theta$) vs. $\theta$ in radians at four radii, $r=\{r_+,
      r_{\rm ISCO}, 10r_g, 10^3r_g\}$, for the GRFFE solution shown in
      Figure~\ref{omegapara9}.  The power is collimated within a half-opening angle of
      $9^\circ$--$14^\circ$ by $r=10^3 r_g$. } \label{energyfluxpara9}
      \end{figure}

  Figure~\ref{energyfluxpara9} shows the angular density of the power
  output.  The peak in the power density occurs at a half-opening
  angle of $9^\circ$--$14^\circ$ at a radius of $r=10^3 r_g$.  The
  integrated electromagnetic power output at radii $r=\{r_+, r_{\rm
  ISCO}, 10r_g, 10^3r_g\}$ is $P = \{0.48, 0.57, 1.39 ,1.77\}$,
  respectively.  Clearly, there is a non-negligible power output from
  the black hole and the transition region between the horizon and the
  ISCO.  There is substantial power from the disk as well.  From
  equation (\ref{avgeff}), the average efficiency is $41\%$ for this
  model, which has a rapidly rotating black hole.  It is slightly
  higher than in the original BZ model for a slowly spinning black
  hole.

\begin{figure}
    \includegraphics[width=3.3in,clip]{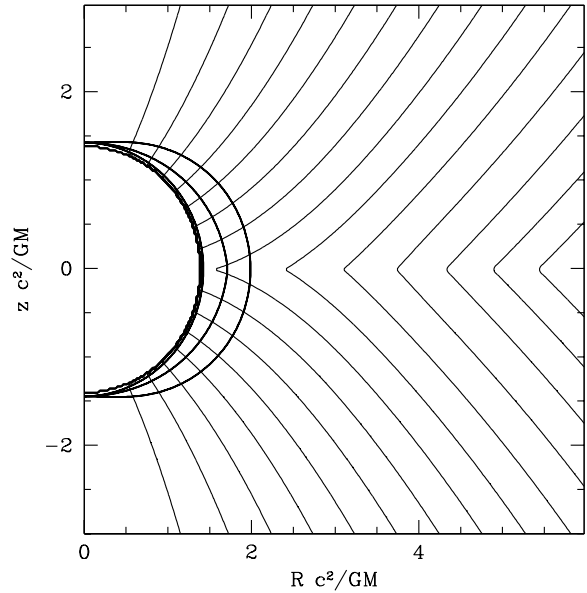}

    \caption{For the paraboloidal ($\nu=1$) model with $a/M=0.9$ and a
    Keplerian disk, this shows the poloidal field geometry given by
    contours of the vector potential ($A_\phi$).  Also shown, from
    inner to outer radius, are the inner-radial computational
    boundary, horizon at $r=1+\sqrt{1-a^2}$, the inner ``light
    cylinder'' corresponding to the \alf~surface for ingoing
    \alf~waves, and the ergosphere at $r=1+\sqrt{1-(a\cos\theta)^2}$.
    Notice that the field geometry becomes monopolar near the horizon
    as in the original Blandford-Znajek solution. } \label{nearbh}
      \end{figure}

  Figure~\ref{nearbh} shows the structure of the magnetosphere once
  the solution has reached a stationary state.  As in the paraboloidal
  solution of Blandford-Znajek, the field geometry becomes monopolar
  near the horizon.  This general relativistic effect becomes
  prominent once the field lines thread the ergosphere.

\begin{figure}
    \includegraphics[width=3.3in,clip]{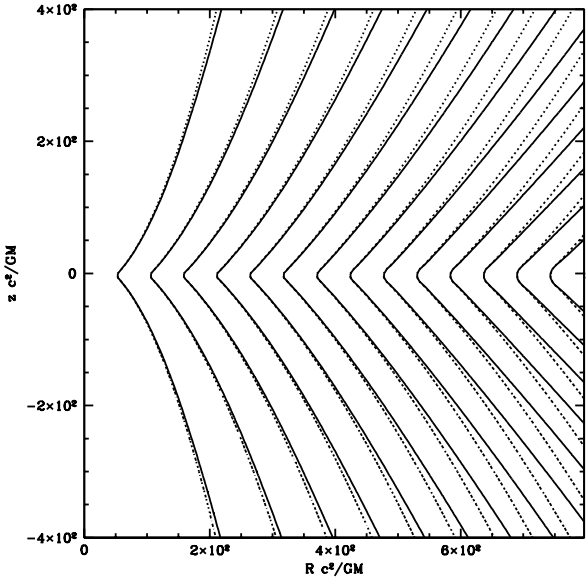}

    \caption{Field lines (constants of $A_\phi$) for the $a/M=0.9$
      paraboloidal ($\nu=1$) model at $t=0$ (initial state,
      non-rotating solution, dotted lines) and $t=1.2\times 10^3t_g$
      (final time, converged rotating solution, solid lines). The
      field is seen to decollimate slightly in the presence of
      rotation, but the effect is quite mild.  Slower and more rapidly
      rotating black holes show similar results. }
      \label{paraa.9_t0_tf} \end{figure}

  Now we consider the radial dependence of certain quantities along a
  field line starting at $\theta\approx 60^\circ$ on the horizon out to
  large radii.  The half-opening angle of the field line is found to
  follow
  \begin{equation}
    \theta_j\approx 54^\circ \left(\frac{r}{2.8r_g}\right)^{-0.48} ,
  \end{equation}
  which is close to the scaling $\theta_j\propto r^{-1/2}$ for the
  non-rotating $\nu=1$ paraboloidal solution \citep{mn06}.  The field
  has apparently slightly decollimated compared to the non-rotating
  case, which is also clear from Figure~\ref{paraa.9_t0_tf} where the
  field lines in the rotating model are seen to be slightly
  decollimated with respect to the nonrotating solution.  Thus it
  appears that the decollimating centrifugal force associated with
  rotation is somewhat stronger than the collimating hoop-stresses
  associated with the toroidal field generated by rotation.

  For a force-free solution, the minimum Lorentz factor with which
  observer at infinity would see particles moving is given by the
  drift speed.  At large radii this minimum Lorentz factor is
  well-fitted by
  \begin{equation}
    \Gamma \equiv u^t \sqrt{-g_{tt}} \approx
    1.5\left(\frac{r}{42.5r_g}\right)^{0.55} ,
  \end{equation}
  which is similar to the scaling $\Gamma\propto r^{1/2}$ found by
  \citet{bn06} for paraboloidal MHD models.  This suggests that the
  force-free flow may be a good model of the acceleration regime in
  GRMHD models, at least at those radii where the magnetic energy is
  much larger than the kinetic energy.

  The orthonormal toroidal field at large radii follows
  \begin{equation}
    B^{\hat{\phi}} \approx B^\phi \sqrt{g_{\phi\phi}} \approx
    0.26\left(\frac{r}{r_g}\right)^{-0.6} ,
  \end{equation}
  and the pitch angle at large radii follows
  \begin{equation}\label{pitchangle}
    \alpha_{\rm pitch} \equiv {\rm
      tan}^{-1}\left(\frac{B^{\hat{r}}}{B^{\hat{\phi}}}\right) \approx
      {\rm tan}^{-1}\left(\frac{c}{\sqrt{g_{\phi\phi}}
        \Omega_F}\right) ,
  \end{equation}
  where the speed of light ($c$) has been temporarily reintroduced.
  This is consistent with an orthonormal radial field of the form
  \begin{equation}
    B^{\hat{r}}\approx -\frac{c}{\Omega_F\sqrt{g_{\phi\phi}}}
    B^{\hat{\phi}} ,
  \end{equation}
  which is a generic feature of force-free solutions with rotating
  field lines (see equation (\ref{equationbtbp}) in
  Appendix~\ref{lorentzscaling}).  Studies of stability suggest that
  force-free fields that follow the above behavior are stable to the
  nonaxisymmetric kink instability \citep{tom01}, though our
  axisymmetric simulations cannot test this.

  \subsection{Summary of Paraboloidal Solutions}

  The original BZ paraboloidal solution could not describe the
  transition region between the accretion disk and black hole, it does
  not account for modification of the field direction for disks with
  Keplerian rotation profiles, and it only applies to slowly spinning
  black holes. Our numerical work has eliminated these restrictions.
  For rapidly rotating black holes, we find an enhanced averaged
  efficiency for the power from the black hole.  We find that power
  output from the transition region between the disk and black hole
  contributes a significant fraction of the total power output from the
  system.  We find that for slowly spinning black holes the transition
  region is important in determining force balance and the power output
  on the horizon.

  \section{GRMHD-Motivated $\nu=3/4$ Solution}\label{fffourth}

  Since GRMHD numerical models produce disk currents consistent with a
  $\nu=3/4$ distribution \citep{mn06}, we now focus on force-free
  solutions with this scaling for the current.  In this section we
  describe the model of the toroidal current and the field rotation
  frequency as deduced from GRMHD simulations, and we study
  corresponding GRFFE numerical models to determine the power and
  collimation of the Poynting-dominated jet and electromagnetic disk
  wind.

  \subsection{Current Model}

  The initial conditions are chosen to be the non-rotating flat-space
  ($M=a/M=0$) $\nu=3/4$ model described in \citet{mn06}.  The vector
  potential is given by
  \begin{equation}
    A_\phi = C r^\nu |\sin\theta| P_{\nu-1}^1(\cos\theta) ,
  \end{equation}
  where $P_l^m$ is the associated Legendre function of the first kind.
  For the case of interest, viz., $\nu=3/4$, the angular solution
  takes the form $P_l^m$ with $l=-1/4$ and $m=1$.  We set $C=1$
  throughout the paper, and this normalizes all energy densities and
  field strengths.  As in the case of the paraboloidal models, this
  vector potential is used to obtain the boundary condition on
  $B^\theta$ at the disk, which is then held fixed for all time.  We
  modify the self-similar solution to account for the transition
  region near the black hole via the mapping given in equation
  (\ref{transB}).  The main effect of this smoothing operation near
  the black hole is the introduction of a split-monopole field near
  the center.  As found in \citet{mg04} and \citet{mckinney2006c}, the
  field is indeed nearly monopolar across the horizon in the
  Poynting-dominated jet region.  This is consistent with the fact
  that all force-free solutions with $M\neq 0$ tend to become
  monopolar near the horizon.

  \subsection{Field Rotation Model}


  We use GRFFE numerical models to investigate the two models of
  $\Omega_F$ described in equations (\ref{model3}) and (\ref{model4}).
  For the latter, we choose $\Omega_{\rm disk} = \Omega_K$, $r_0=2
  r_+$ and $r_{\rm trans\Omega}=3r_+$, though we have found that
  choosing $r_0=r_+$ and $r_{\rm trans\Omega}=2 r_+$ leads to only
  small quantitative differences.  To fit the time-averaged GRMHD
  numerical model shown in Figure 8 in \citet{mn06}, one could choose
  $\Omega_{\rm DBH} = 0.6 \Omega_H$.  For general $a/M$, a good fit is
  provided by $\Omega_{\rm DBH}\sim\Omega_H/2$ for $a/M\gtrsim 0.4$
  and $\Omega_{\rm DBH} = \Omega_H$ for slower black hole spins
  \citep{mn06}.  Note that $\Omega_F$ is the same in Kerr-Schild and
  Boyer-Lindquist coordinates, and $\Omega$ is also the same at the
  equator.  Thus, the above prescription for $\Omega_F$ in the
  equatorial disk is valid for both coordinate systems.

  \subsection{Rotating Black Hole and a $\nu=3/4$ Keplerian Disk}\label{fiducialmodel}

  We consider a GRFFE model with a black hole spin of $a/M=0.9$ and an
  equatorial current sheet with a $\nu=3/4$ power law current
  distribution.  Our fiducial model for the transition region, called
  ``TModel 1,'' smooths the field vector potential using equation
  (\ref{transB}) with $r_{\rm transB}=2r_+$, and models the field
  rotation via equation (\ref{model3}) with $\Omega_{\rm DBH}/\Omega_H
  = 0.32$ and $r_{\rm trans\Omega}=2r_+$.

  For this transition model we seek to find force-free solutions with
  no discontinuities.  If one chooses $\Omega_{\rm DBH} = 0.6
  \Omega_H$ as in the GRMHD simulations, discontinuities appear above
  the equatorial plane.  This is expected since the disk is not
  force-free.  The value $\Omega_{\rm DBH} = 0.32 \Omega_H$ was chosen
  to give a smooth transition from the disk to the black hole so that
  there are negligible discontinuities near the accretion disk - black
  hole interface that would correspond to extra toroidal currents in
  the magnetosphere.  Other transition models are considered in
  section~\ref{deptrans}, such as one where we do choose $\Omega_{\rm
  DBH} = 0.6 \Omega_H$ and allow discontinuities to be present in the
  magnetosphere.

\begin{figure}
    \includegraphics[width=3.3in,clip]{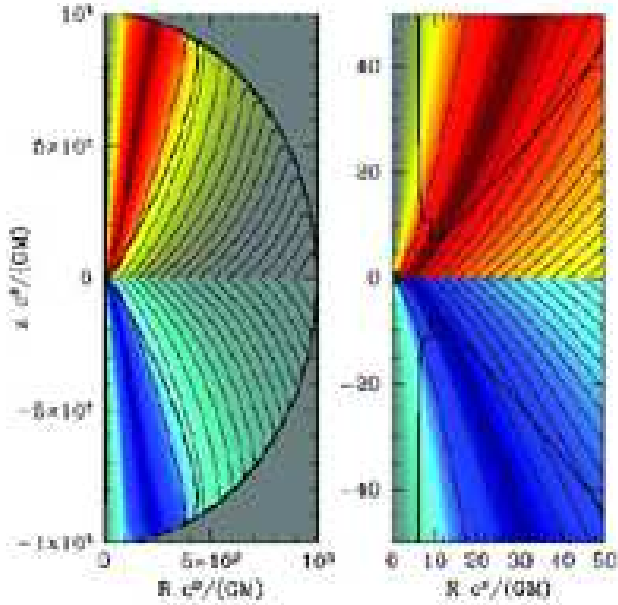}

    \caption{Field lines (black lines), enclosed poloidal current
      ($B_\phi$, where $B_\phi\approx
      B^{\hat{\phi}}\sqrt{g_{\phi\phi}}$, red positive, blue
      negative), and light ``cylinder'' (thick black line).  The
      results correspond to a GRFFE model with $\nu=3/4$ at time
      $t=1200t_g$ for a rotating disk around a spinning black hole
      with $a/M=0.9$.  The panel on the left shows the full
      computational domain and that on the right shows a closeup of
      the central region.  The black hole and the disk together
      generate a Poynting outflow in the form of a jet surrounded by a
      wind, with power focused along the polar axis. Small artifacts
      at the outer radial edge in the left panel are the result of
      interpolation and reflections of the Poynting jet off the outer
      radial boundary. } \label{biffdeplot} \end{figure}

  Starting with the non-rotating force-free solution as initial
  conditions, the above model was simulated with the GRFFE code until it
  reached steady state ($t=1200t_g$).  Figure~\ref{biffdeplot} shows the
  poloidal structure of field lines (also lines of constant $A_\phi$),
  the enclosed poloidal current from the pole ($B_\phi$), and the light
  ``cylinders'' (only the outer light cylinder is clearly visible) in
  the final state.  The poloidal field geometry has changed little from
  the initial solution, showing that rotation has a negligible effect.
  The polar enclosed poloidal current ($B_\phi$) is shown in color,
  where the radial energy flux is given by $F^r_E = \Omega_F B_\phi
  B^r$.  Note that for stationary, axisymmetric flows $B_\phi$ and
  $\Omega_F$ are constant along field lines.

  The field lines in the $\nu=3/4$ solution show less poloidal
  collimation than the paraboloidal solution but more than the
  monopole solution (which has no collimation at all).  As compared to
  the GRMHD solution, the pressure support of the corona has been
  replaced by the stable force-free field lines, but otherwise the
  funnel region is quantitatively similar to the GRMHD solution.

\begin{figure}
    \includegraphics[width=3.3in,clip]{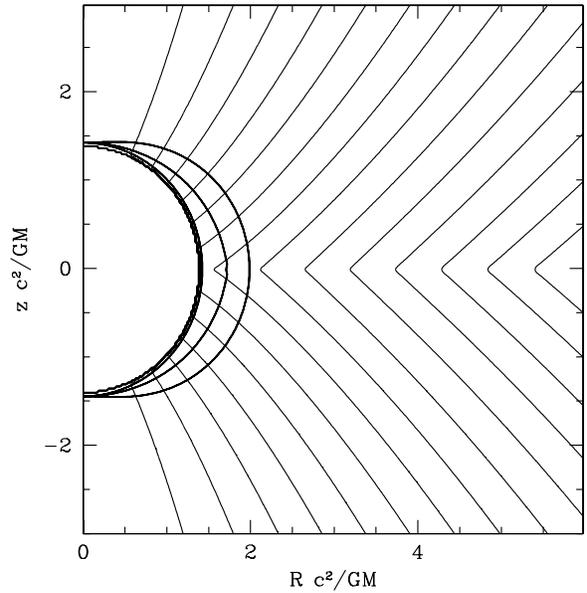}

    \caption{Similar to Figure~\ref{nearbh}, but for the
    GRMHD-motivated, $a/M=0.9$, Keplerian disk model with a $\nu=3/4$
    current distribution. Notice that the field geometry becomes
    slightly monopolar near the horizon. } \label{nearbhfid}
      \end{figure}

  Figure~\ref{nearbhfid} shows the structure of the magnetosphere once
  the solution has reached a stationary state.  As in the paraboloidal
  solution described previously, the field geometry becomes slightly
  monopolar near the horizon.  This general relativistic effect
  becomes important once the field lines thread the ergosphere.

\begin{figure}
    \includegraphics[width=3.3in,clip]{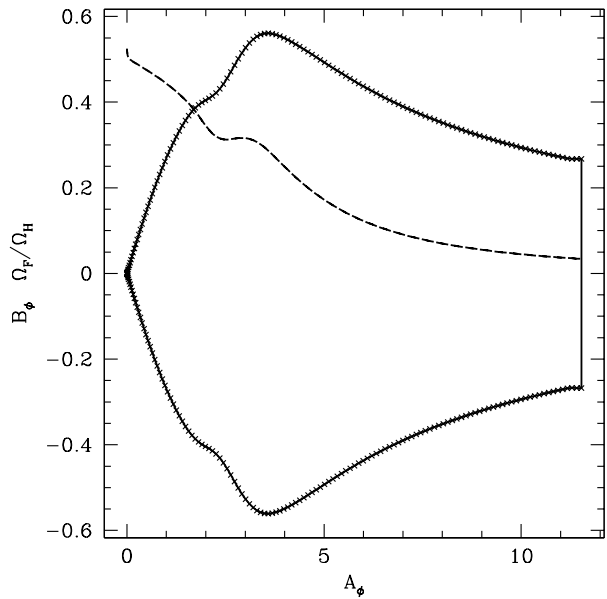}

    \caption{Enclosed poloidal current ($B_\phi$, where $B_\phi\approx
      B^{\hat{\phi}}\sqrt{g_{\phi\phi}}$, solid line with crosses) and
      field angular frequency per unit black hole angular frequency
      ($\Omega_F/\Omega_H$, dashed line) vs. the vector potential
      ($A_\phi=\Psi$, also called the flux function) at spherical
      polar radius $r=20$.  This shows that the poloidal current
      increases in the jet and then decreases at larger radii in the
      disk that contributes to a line return current.  There is no
      enclosed current at $r=\infty$ unlike in the paraboloidal case.
      This also shows how the rotation frequency of the field lines
      behaves near the black hole and how it connects onto the
      Keplerian disk at large radii (large $A_\phi$). }
      \label{bphiaphi}
      \end{figure}

  Figure~\ref{bphiaphi} diagnoses the poloidal current flow and
  angular rotation of field lines in the force-free magnetosphere at a
  radius of $r=20$.  The value of $B_\phi$ is the amount of poloidal
  current enclosed away from the polar axis down to the given value of
  $A_\phi$, where $A_\phi=0$ on the polar axis.  This plot shows that
  there is an increase in current away from the polar axis due to the
  jet coming from the black hole, and then there is a drop in current
  as the disk supplies the return current.  At large radii (large
  $A_\phi$) the current vanishes.  The current closes at $r=0$ and
  $r=\infty$ in a non-singular manner, unlike in the paraboloidal case
  where $B_\phi$ is finite at large radii so that the radially
  integrated poloidal current is infinite. The small-scale oscillation
  in $B_\phi$ is located at the black hole jet-disk interface, and the
  detailed behavior of this region depends on the black hole spin and
  disk rotation profile. This current structure is comparable to that
  of neutron star magnetospheres (see, e.g. \citealt{mckinney2006b}).
  The plot also shows $\Omega_F/\Omega_H$ as a function of $A_\phi$.
  This shows how the black hole solution at small $A_\phi$ matches
  onto the Keplerian disk at large $A_\phi$.

  \subsubsection{Angular Dependence within Jet}

  \begin{figure}
    \includegraphics[width=3.3in,clip]{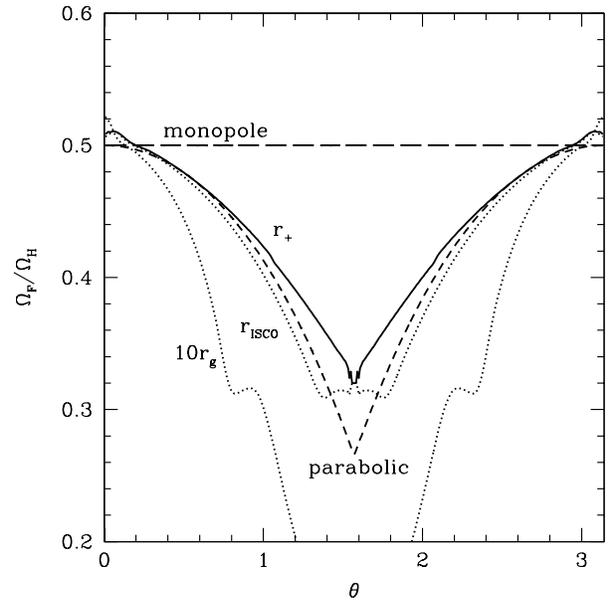}

    \caption{Field angular frequency per unit black hole angular frequency
      ($\Omega_F/\Omega_H$) vs. $\theta$ in radians on the horizon ($r=r_g$, solid line) and
      at two other radii ($r=r_{\rm ISCO}\approx 2.3209r_g$ and
      $r=10r_g$, dotted lines).  The results correspond to a GRFFE
      model with $\nu=3/4$ at time $t=1200t_g$ for a rotating disk
      around a spinning black hole with $a/M=0.9$.  Also shown are the
      Blandford-Znajek monopole solution (long-dashed line) and
      paraboloidal solution (short-dashed line) on the horizon. Note
      that the field angular frequency at the horizon is roughly
      similar for the paraboloidal ($\nu=1$) and $\nu=3/4$ models. }
      \label{omegaffourth} \end{figure}

  Figure~\ref{omegaffourth} shows $\Omega_F/\Omega_H$ for the
  $\nu=3/4$ solution at $t=1200t_g$ at three different radii and also
  shows the monopole and paraboloidal solutions on the horizon.  As
  expected, the $\nu=3/4$ solution has a field rotation at the horizon
  ($\Omega_F/\Omega_H\approx 0.32$) that is close to the paraboloidal
  solution ($\nu=1$, $\Omega_F/\Omega_H\approx 0.32$ for $a/M=0.9$)
  and lower than the monopole solution ($\nu=0$,
  $\Omega_F/\Omega_H\approx 0.5$).  The profile of $\Omega_F/\Omega_H$
  is similar to the GRMHD models shown in Figure 8 (left panel) in
  \citet{mg04}.  In their plot the value of $\Omega_F/\Omega_H\approx
  0.32$ near the transition between the force-free region and the
  accretion disk region.  This is further evidence that the $\nu=3/4$
  GRFFE model gives a solution that is consistent with the
  Poynting-dominated jet found in full GRMHD. Close to the poles
  $\Omega_F/\Omega_H$ deviates from the expected value of
  $\Omega_F/\Omega_H\approx 1/2$ due to the coordinate singularity.
  However, this feature has negligible impact on the solution at large
  radii due to the collimation of field lines toward the polar axis.

  \begin{figure}
    \includegraphics[width=3.3in,clip]{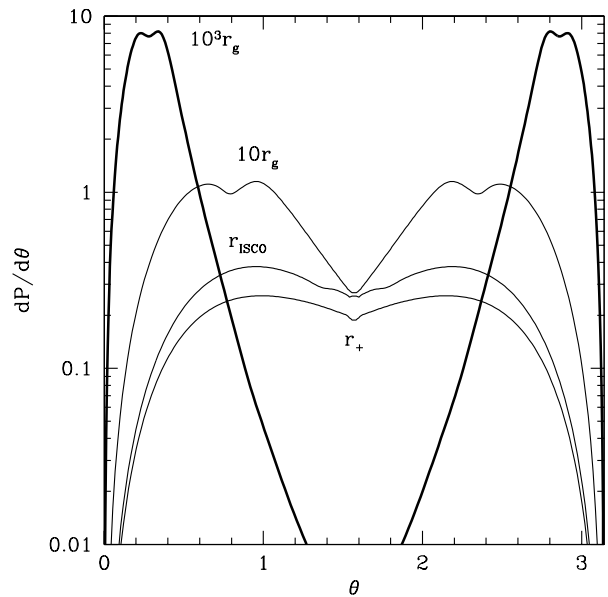}

    \caption{Angular density of the electromagnetic power output
      ($dP/d\theta$) vs. $\theta$ in radians at four radii, $r=\{r_+,
      r_{\rm ISCO}, 10r_g, 10^3r_g\}$, for the GRFFE solution shown in
      Figure~\ref{omegaffourth}. The power is seen to be collimated within a
      half-opening angle of $15^\circ$--$26^\circ$ by $r=10^3 r_g$.
      The collimation is significantly less than in the otherwise
      similar paraboloidal ($\nu=1$) model. } \label{energyflux}
      \end{figure}

  Figure~\ref{energyflux} shows the angular density of the power
  output. The $\nu=3/4$ disk provides most of its energy inside
  $r=100r_g$. The integrated electromagnetic power output at $r=\{r_+,
  r_{\rm ISCO}, 10r_g, 10^3r_g\}$ is $P = \{0.47, 0.62, 1.50 ,1.76\}$,
  respectively. The figure shows that the peak in the power is
  collimated to within a half-opening angle of $15^\circ$--$26^\circ$ by
  $r=10^3 r_g$.  Using equation (\ref{avgeff}), we calculate an average
  efficiency of $41\%$ for this model, which is the same as the
  equivalent paraboloidal model.

  One can compare the electromagnetic power outputs in the GRFFE model
  to those found in the equivalent GRMHD model discussed in
  \citet{mn06}.  The powers are comparable at the horizon.  However, in
  the GRMHD model, the electromagnetic power from the disk is lost to
  the matter and only the power from the black hole survives at large
  radii, whereas in the GRFFE model all the electromagnetic power output
  of the black hole and the disk reaches large radii.

  \subsubsection{Radial Dependence within Jet}

  \begin{figure*}

    \begin{center}
      \includegraphics[width=6.6in,clip]{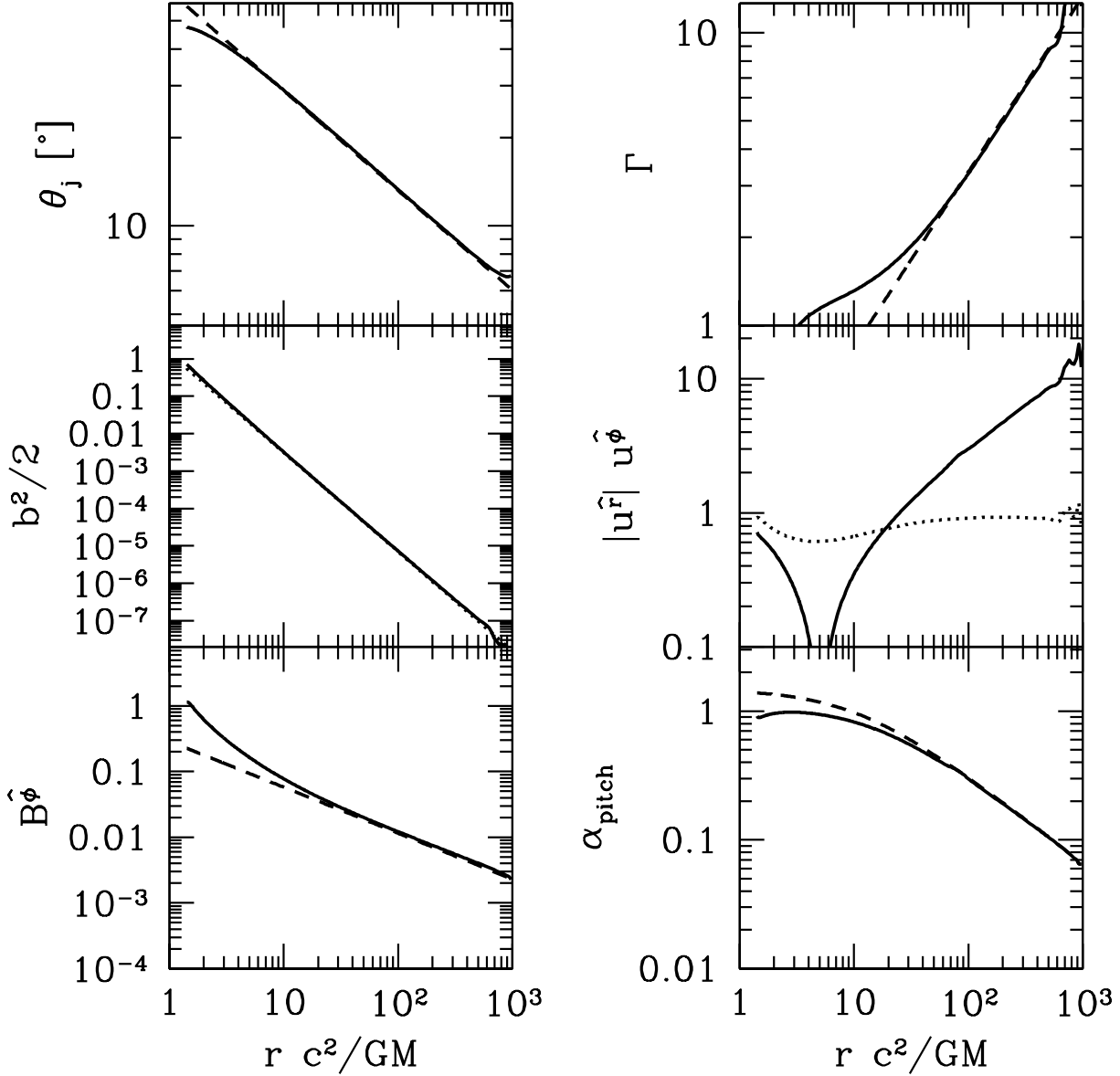}
    \end{center}

    \caption{Variation with radius of various quantities in the
      converged GRFFE solution with $\nu=3/4$ and a rotating disk
      around a spinning black hole with $a/M=0.9$.  The solid lines in
      the panels show the half-opening angle of the jet in degrees
      ($\theta_j$), the Lorentz factor ($\Gamma$), the comoving energy
      density ($b^2/2$), the orthonormal radial velocity
      ($u^{\hat{r}}$), the orthonormal toroidal field strength
      $(B^{\hat{\phi}})$, and the pitch angle in radians ($\alpha_{\rm
      pitch}$).  The features near the outer radial boundary are due
      to reflections of initial transients off the boundary.  Except
      for the panel with $\alpha_{\rm pitch}$, the dashed lines in the
      other panels show power-law fits to the numerical results.  The
      dotted line in the $u^{\hat{r}}$ panel shows the orthonormal
      angular velocity $u^{\hat\phi}$.  The dashed line in the
      $\alpha_{\rm pitch}$ panel shows the limit below which the kink
      instability will be present according to \citet{tom01}.  It
      indicates that the flow is marginally stable at large radii.
      The other panels show that the solution has a power-law behavior
      at large radii. } \label{jetstructgrffe}
      \end{figure*}

  Figure~\ref{jetstructgrffe} shows the jet half-opening angle
  ($\theta_j$), the Lorentz factor ($\Gamma\equiv u^t
  \sqrt{-g_{tt}}$), the comoving energy density ($b^2/2$), the
  approximate orthonormal absolute value of radial and $\phi$
  velocities ($|u^{\hat{r}}|\equiv |u^r| \sqrt{g_{rr}}$ and
  $u^{\hat{\phi}}\equiv u^\phi \sqrt{g_{\phi\phi}}$), the orthonormal
  toroidal field strength ($B^{\hat{\phi}}\equiv B^\phi
  \sqrt{g_{\phi\phi}}$), and the pitch angle within the jet
  ($\alpha_{\rm pitch} \equiv {\rm
    tan}^{-1}(B^{\hat{r}}/B^{\hat{\phi}})$).  The figure also shows
  power-law fits to some quantities.  This figure can be compared to
  panels in figures 7 and 8 in \citet{mckinney2006c}.

  The opening angle of the jet for the field line starting at
  $\theta_j\approx 57^\circ$ has a power-law dependence at large radii
  that follows
  \begin{equation}\label{thetajnu34}
  \theta_j\approx 57^\circ \left(\frac{r}{2.8r_g}\right)^{-0.34} ,
  \end{equation}
  compared to $\theta_j\propto r^{-0.375}$ for the non-rotating
  $\nu=3/4$ solution \citep{mn06}.  This shows that the field has
  slightly decollimated compared to the non-rotating solution, but the
  effect is weak.  The power-law dependence for $\theta_j$ is
  strikingly similar to that found for the opening angle of the jet in
  GRMHD simulations given by equations (24) and (25) in
  \citet{mckinney2006c}.  For example, the choice of $\theta_j\approx
  57^\circ$ on the horizon for the field line corresponds to where the
  Poynting-dominated jet starts in the GRMHD simulations of
  \citet{mckinney2006c} and corresponds to where the combined
  disk+corona scale height to radius ratio is $H/R\approx 0.6$.  In
  this force-free model, $\theta_j\approx 5^\circ$ at $r\approx
  5\times 10^3r_g$, a similar value as found in \citet{mckinney2006c}
  for the core of the jet at large radii.

  The GRMHD simulations and the present force-free model have an
  opening angle that is consistent up to the point where in the GRMHD
  model the jet becomes conical due to dissipative processes.  In the
  GRMHD simulations the power-law dependence slightly depends on the
  field line chosen, while in these GRFFE simulations the power-law
  dependence is the same for all field lines.

  The Lorentz factor at large radii is well-fitted by
  \begin{equation}
    \Gamma \approx 2\left(\frac{r}{42.5r_g}\right)^{0.6} ,
  \end{equation}
  which is consistent with the magnetic acceleration domain found in
  \citet{mckinney2006c}.  This behavior is also similar to the analytic
  result of $\Gamma\propto r^{1/2}$ obtained by \citet{bn06} for
  paraboloidal MHD winds.

  As in the GRMHD model of \citet{mckinney2006c}, the plasma angular
  velocity ($u^{\hat{\phi}}$) remains subrelativistic while the radial
  velocity ($u^{\hat{r}}$) becomes relativistic.

  The comoving energy density follows
  \begin{equation}
    \frac{b^2}{2} \approx 1.4\left(\frac{r}{r_g}\right)^{-2.65} ,
  \end{equation}
  which is close to the scaling $b^2\propto r^{-2.5}$ found for the
  non-rotating $\nu=3/4$ solution.  This implies that the black hole and
  disk rotation has only a small effect on the field energy.

  The orthonormal toroidal field follows
  \begin{equation}
    B^{\hat{\phi}} \approx 0.29\left(\frac{r}{r_g}\right)^{-0.7} ,
  \end{equation}
  which is remarkably close to the GRMHD solution found in
  \citet{mckinney2006c} for the inner-radial region before dissipation
  becomes important at $r\sim 10^3r_g$.

  The pitch angle is found to follow
  \begin{equation}
    \alpha_{\rm pitch} \approx {\rm tan}^{-1}\left(\frac{c}{g_{\phi\phi}
      \Omega_F}\right) ,
  \end{equation}
  with the speed of light ($c$) temporarily reintroduced.  As for the
  paraboloidal models, the orthonormal radial field at large radii is
  well-fitted by
  \begin{equation}
    B^{\hat{r}}\approx -\frac{c}{\Omega_F\sqrt{g_{\phi\phi}}} B^{\hat{\phi}} .
  \end{equation}

  Summarizing this section, there are two main results.  First, the
  poloidal structure of the field in the rotating GRFFE models
  considered in this paper is rather similar to the field structure
  found in the non-rotating models discussed in \citet{mn06}.  Since
  we showed previously that the latter models agree well with the
  poloidal field structure in the jet region of GRMHD models, the
  present models also agree well.  Second, the structure of the
  toroidal field, along with various related quantities such as the
  pitch angle, Poynting flux, jet acceleration, Lorentz factor, etc.,
  obtained from the rotating GRFFE solution agree very well with the
  corresponding quantities in the GRMHD jet.  Thus, the rotating GRFFE
  model appears to be an excellent representation of all facets of
  jets.

  \subsubsection{Stability}

  The pitch angle of the magnetic field lines becomes progressively
  smaller at large distances, becoming as small as $\alpha_{\rm
  pitch}\approx 3^\circ$ at $r=10^3 r_g$.  This is consistent with the
  Poynting-dominated jet studied in \citet{mckinney2006c}.  The small
  pitch-angle argues that the flow might be kink unstable, since the
  Kruskal-Shafranov criterion for instability is abundantly satisfied.
  The criterion for instability is given by
  \begin{equation}
    \left|\frac{B^{\hat{\phi}}}{B^{\hat{r}}}\right| \gtrsim \frac{R}{L} ,
  \end{equation}
  where $L\approx \sqrt{g_{\theta\theta}}$ is approximately the vertical
  length of the jet and $R\approx \sqrt{g_{\phi\phi}}$ is approximately
  the cylindrical radial extent of the jet.

  \citet{tom01} considered the effect of rotation on stability and found
  that a Poynting-dominated jet is kink unstable only if both the
  Kruskal-Shafranov criterion {\it and} the additional condition
  \begin{equation}
    \left|\frac{B^{\hat{\phi}}}{B^{\hat{r}}}\right| \gtrsim \frac{R}{R_L} ,
  \end{equation}
  is also satisfied, where $R_L=c/\Omega_F$ is the light cylinder
  radius \citep{tom01}.  Their result only strictly applies inside the
  light cylinder and for a uniform field, but it does suggest that
  rotation could stabilize a jet that might appear unstable according
  to the Kruskal-Shafranov criterion.  The lower right panel of
  Figure~\ref{jetstructgrffe} compares the quantity ${\rm tan}^{-1}
  (R_L/\sqrt{g_{\phi\phi}})$ with the pitch angle, and suggests that
  the flow is marginally stable to the kink instability at large
  radii.  Note, however, that the kink instability is
  non-axisymmetric, whereas our simulations are axisymmetric.  Thus,
  3D simulations are needed to check the \citet{tom01} stability
  analysis.

  We can, however, study the stability of our numerical force-free
  solutions to the {\it axisymmetric} pinch instability both inside
  and outside the light cylinder.  For $a/M=0.9$, the light cylinder
  is at $R\approx 6r_g$--$10r_g$. For the fiducial field line that
  threads the black hole (as given by equation (\ref{thetajnu34}), the
  outer value of the cylindrical radius is $R\approx 120r_g$.  Thus
  such a field line is far beyond the light cylinder.  Field lines
  that thread the outer disk have not yet passed through their light
  cylinder, so we focus on the stability of the solution close to the
  black hole.

  Once the solution became stationary by $t\sim 10^3 t_g$, random
  perturbations with an amplitude of $10\%$ were added.  The
  Poynting-jet from the black hole and the electromagnetic wind from
  the disk were found to be stable to these perturbations within and
  far beyond the light cylinder where the toroidal field dominated.
  Thus, we can state that the axisymmetric pinch mode does not grow,
  at least in an Eulerian sense (though it may still grow in a
  Lagrangian sense moving out with the fluid and may manifest itself
  at large radii).

  \subsection{Dependence on Black Hole Spin}\label{bhspin}

  Now we study the dependence of the power on the black hole spin for
  GRMHD-motivated GRFFE models.  In the GRMHD simulations, the value
  of $\Omega_F/\Omega_H$ in the disk depends strongly on whether the
  spin is larger or smaller than $a/M\sim 0.4$, thus a mapping to a
  force-free model is not straight-forward because such force-free
  models choose an approximately fixed value of $\Omega_F/\Omega_H$
  for the entire magnetosphere above the equatorial plane. As
  described in \citet{mn06}, GRMHD models with black hole spins
  $a/M\gtrsim 0.4$ tend to have $\Omega_F/\Omega_H\sim 1/2$ on the
  horizon, and so using $\Omega_F/\Omega_H=0.32$ in the GRFFE modeling
  should lead to a solution close to the GRMHD solution.  However, for
  $a/M\lesssim 0.4$, the GRMHD solution within the disk is found to
  follow $\Omega_F/\Omega_H\sim 1$ because the disk dominates the
  field all the way through the horizon.  Since the power from the
  black hole is $P\propto \Omega_F (\Omega_F-\Omega_H) (B^r)^2$, then
  $P\sim 0$ for $a/M\lesssim 0.4$.

  We already know that the magnetosphere is free from unwanted
  discontinuities only if one chooses $\Omega_F/\Omega_H\sim 0.32$ on
  the horizon at the equator.  One problem with modelling $a/M\lesssim
  0.4$ black holes is that choosing the value of
  $\Omega_F/\Omega_H\sim 1$ would lead to discontinuities in the
  magnetosphere.  Even if one allowed discontinuities, the more
  important problem is that the field lines passing through the
  horizon above the equator would still settle on having
  $\Omega_F/\Omega_H=0.32$.  Hence, while the power through the disk
  would vanish exactly at the equator, it would not vanish just off
  the equator.  A comparable GRMHD simulation with a {\it thick} disk
  would have $P\sim 0$ over the {\it entire} part of the horizon
  occupied by the infalling disk material.  In such models, as the
  disk thickness increases, the power (per unit $(B^r)^2$) drops
  because the disk takes over the magnetosphere.  Thus, the disk
  thickness plays an important role in setting the total power output
  for $a/M\lesssim 0.4$.

  Despite these complications, in the following we employ the {\it same
    model} of $\Omega_F/\Omega_H$ in the disk for all black hole spins,
  but we primarily focus on the models with $a/M\gtrsim 0.5$ where the
  results are most representative of the GRMHD models for any disk
  thickness.  We use equation (\ref{model3}) to model the disk rotation
  and equation (\ref{transB}) to soften the vector potential in the
  transition region.

\begin{table*}
\begin{center}
\begin{tabular}{llllllll}
\hline
\multicolumn{8}{|c|}{\bfseries Black Hole Spin Study} \\
\hline
$a$ & $P$ & $P$ & $P$  & $P$  & $\Gamma$ & $\theta_j$ & $\theta_j$ \\
  & $[r_+]$ & $[r_{\rm ISCO}]$ & $[10r_g]$  & $[10^3r_g]$  & $[10^3r_g]$ & $[10^3r_g]$ & $[5\times 10^3r_g]$ \\
  &   &  &  &  &  &[{\rm Peak Power}] & $[H/R=0.6]$  \\

\hline
\multicolumn{8}{|c|}{\bfseries Normalized by Disk Field Strength} \\
\hline

0.1 & 0.02415  &    0.5734   &    1.795  &     3.125 & 7 & $25^\circ$
  & $4^\circ$ \\

0.2 & 0.1122   &    1.283   &    2.694  &     3.995 & 8 & $25^\circ$& $4^\circ$  \\

0.5 & 0.7368   &    1.258   &    3.109  &      4.41 & 8 &  $15^\circ$--$26^\circ$& $4^\circ$  \\

0.8 & 1.893  &     2.632 &       5.721   &     6.92 & 10 &   $15^\circ$--$26^\circ$& $4^\circ$  \\

0.9 &  2.097  &     2.744  &     6.662  &     7.846 & 15 &
$12^\circ$--$26^\circ$
  & $5^\circ$ \\

0.9375 & 2.016 &      2.552   &    6.889  &     8.187 & 20 &
  $12^\circ$--$26^\circ$ & $5^\circ$ \\

\\
\hline
\multicolumn{8}{|c|}{\bfseries Normalized by Black Hole Field Strength} \\
\hline

0.1 & 0.01562  &    0.3709 &      1.161  &     2.022 & 7 & $25^\circ$ & $4^\circ$ \\

0.2 & 0.05428  &    0.6207 &      1.304  &     1.933 & 8 & $25^\circ$ & $4^\circ$ \\

0.5 & 0.3417   &   0.5835  &     1.442   &    2.045  & 8 &  $15^\circ$--$26^\circ$& $4^\circ$  \\

0.8 & 0.5478   &   0.7618  &     1.656    &   2.003 & 10 &  $15^\circ$--$26^\circ$ & $4^\circ$ \\

0.9 & 0.4715   &   0.6168   &    1.498   &    1.764 & 15 &  $12^\circ$--$26^\circ$
  & $5^\circ$ \\

0.9375 & 0.3905  &     0.4942  &      1.334  &      1.586 & 20 &
  $12^\circ$--$26^\circ$ & $5^\circ$ \\

\hline
\end{tabular}
\caption{Electromagnetic Power, Lorentz factor, and half-opening angle of
jet for different black hole spins and at a few radii for each model. The
second to last column shows $\theta_j$ at the location of the peak in the
angular power, which includes a significant power from disk at low black
hole spins.  The last column shows $\theta_j$ at a large radius for the
field line that starts at $\theta_j=57^\circ$ on the horizon, which begins
the force-free region in GRMHD numerical models that have an accretion disk
up to $H/R\approx 0.3$ and a corona up to $H/R\approx 0.6$. } \label{tbl2}
\end{center}
\end{table*}

  Table~\ref{tbl2} shows the electromagnetic power output, Lorentz
  factor, and half-opening angle for different black hole spins and at
  different radii.  The power could be arbitrarily normalized, but we
  use two specific normalizations that are convenient for comparisons
  with the results from GRMHD models.  In the upper section of the
  table, the power has been normalized by $(B^r)^2$ on the horizon at
  the poles of the {\it initial} $a/M=M=0$ model with $\nu=3/4$.  Since
  the field strength at large radius is not significantly changed by the
  slow disk rotation or the rotation of the black hole, this is
  equivalent to a normalization by the {\it disk field strength at large
    radii}.  If the accretion rate and field strength at large radii can
  be established for a GRMHD model, then this is a good normalization to
  use for comparison with the GRFFE model.  The lower section of the
  table has the power normalized by $(B^r)^2$ on the horizon at the
  poles of the {\it final} stationary model.  This removes the changes
  in the field strength near the black hole due to the spin of the black
  hole.  If the accretion rate at small radii is in force-balance with
  the field near the black hole, then this is a good
  normalization. However, the field strength near the horizon is a
  strong function of black hole spin \citep{mckinney2005a} and a mild
  function of the mass accretion rate, so this normalization should be
  considered carefully when comparing results for different black hole
  spins.

  Regardless of the normalization, one interesting conclusion, from
  the table values for $P[r_+]$ to $P[r=10^3r_g]$, is that even at
  high black hole spins the disk provides no less than $\approx 1/4$
  of the total power output.

  The Lorentz factor and $\theta_j$ are independent of the normalization
  of course.  The power shifts to small angles due to the emergence of
  the black hole component above $a/M\sim 0.5$, although this black hole
  component is always present.  For the $a/M=0.9$ model, the field line
  that connects to the black hole at the equator eventually reaches
  $\theta_j\approx 10^\circ$ at $r=10^3r_g$.  The peak in the angular
  power density is set by the black hole power output for $a/M\gtrsim
  0.9$ and is otherwise set by the disk power output.

  If a disk were present with a sharply defined height to radius ratio
  of $H/R\approx 0.3$, as in the GRMHD numerical models, and the matter
  did not modify the poloidal field direction (which is a good
  approximation; see \citealt{mn06}), then the field line would go from
  $\theta\approx 72^\circ$ on the horizon to $\theta_j\approx 10^\circ$
  at $r=10^3r_g$.  In the presence of both an accretion disk and corona,
  as in the GRMHD numerical models, with a combined scale-height to
  radius ratio of $H/R\approx 0.6$, then $\theta_j\approx 7^\circ$ at
  $r=10^3r_g$.  Finally, if we extend the calculation to $r=5\times
  10^3r_g$ and again consider a disk+corona region blocking part of the
  power of the jet, then $\theta_j\approx 5^\circ$ there.  This is the
  same jet collimation found for the core of the jet in time-dependent
  GRMHD numerical models \citep{mckinney2006c}.  In those simulations
  the jet had an extended low-energy wing out to $\theta_j\approx
  27^\circ$, which could be due to the effects of matter or due to a
  lack of transfield pressure support at large radii.  Overall the
  opening angle of the jet and the power output measured in the GRMHD
  numerical models and the $\nu=3/4$ force-free models are in excellent
  agreement.  A monopolar, paraboloidal, or cylindrical force-free model
  would not agree this well.

\begin{figure}
    \includegraphics[width=3.3in,clip]{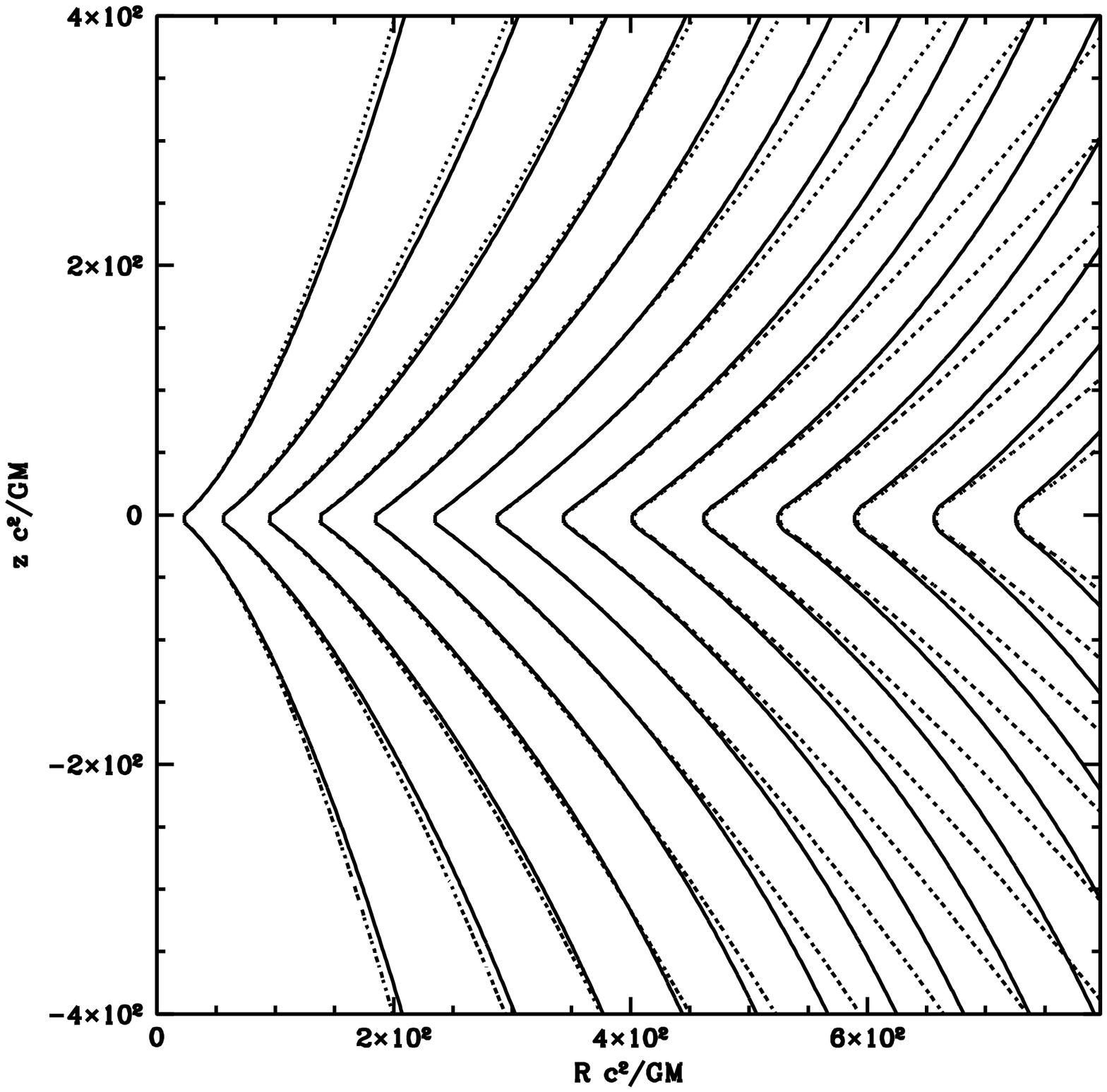}

    \caption{Field lines for the $a/M=0.1$ GRFFE model with $\nu=3/4$
      at $t=0$ (initial state, nonrotating solution, dotted lines) and
  $t=1.2\times 10^3t_g$ (final converged rotating solution, solid
  lines). The field lines threading the black hole show mild
  decollimation, as in the paraboloidal case, and the field lines from
  the outer regions of the disk show some collimation. }
  \label{a.1_t0_tf} \end{figure}

\begin{figure}
    \includegraphics[width=3.3in,clip]{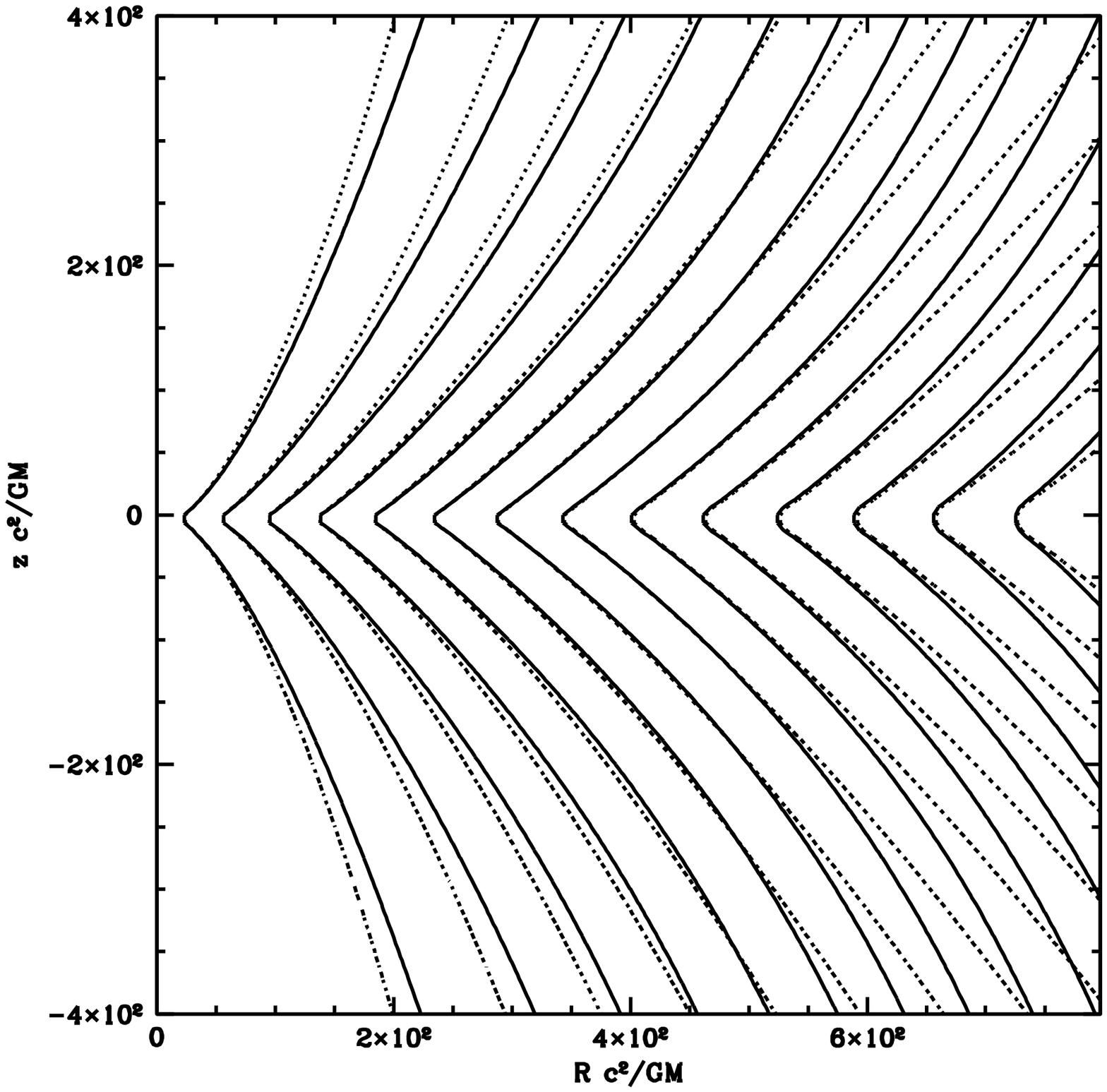}

    \caption{Similar to Figure~\ref{a.1_t0_tf}, but for a rapidly spinning black hole
      with $a/M=0.9375$.  There is again mild decollimation of the
      field lines threading the black hole, and some collimation of
      field lines from the disk. }
  \label{a.9375_t0_tf} \end{figure}

  Figures~\ref{a.1_t0_tf} and~\ref{a.9375_t0_tf} show the initial and
  final configurations of the field lines for the $a/M=0.1$ and
  $a/M=0.9375$ models, respectively.  We see that the Keplerian rotation
  of the disk leads to some collimation of the disk field lines. On the
  other hand, the spin of the black hole leads to modest {\it
    decollimation} of the field lines near the poles.

  \subsection{Dependence on the Transition Model}\label{deptrans}

  Different models of the transition region were studied and
  Table~\ref{tbl3} gives the same information as Table~\ref{tbl2} for
  these different models.

  The first model, called ``TModel 1,'' has been already discussed in
  \S\S~\ref{fiducialmodel} and~\ref{bhspin}.  It uses equation
  (\ref{model3}) for $\Omega_F$ with $\Omega_{\rm DBH}/\Omega_H =
  0.32$ and $r_{\rm trans\Omega}=2r_+$, and equation (\ref{transB})
  for the vector potential with $r_{\rm transB}=2r_+$.

\begin{table*}
\begin{center}
\begin{tabular}{llllllll}
\hline
\multicolumn{8}{|c|}{\bfseries Transition Model Study} \\
\hline
Model Type & $P$ & $P$ & $P$  & $P$  & $\Gamma$ & $\theta_j$ & $\theta_j$ \\
  & $[r_+]$ & $[r_{\rm ISCO}]$ & $[10r_g]$  & $[10^3r_g]$  & $[10^3r_g]$ & $[10^3r_g]$ & $[5\times 10^3r_g]$ \\
  &   &  &  &  &  &[{\rm Peak Power}] & $[H/R=0.6]$  \\

\hline
\multicolumn{8}{|c|}{\bfseries Normalized by Disk Field Strength} \\
\hline

{\bfseries TModel 1} & 2.097 & 2.744 & 6.662 & 7.846 & 15 &  $12^\circ$--$26^\circ$ & $5^\circ$ \\

{\bfseries TModel 2} &  1.672 & 1.938 & 3.983 & 4.65 &  15 & $15^\circ$--$26^\circ$  & $5^\circ$ \\

{\bfseries TModel 3} & 10.64 & 11.47 & 20.82 & 21.84 & 15 &  $12^\circ$--$26^\circ$ & $5^\circ$\\

\\
\hline
\multicolumn{8}{|c|}{\bfseries Normalized by Black Hole Field Strength} \\
\hline

{\bfseries TModel 1} & 0.4715   &   0.6168   &    1.498   &    1.764 & 15 &  $12^\circ$--$26^\circ$ & $5^\circ$ \\

{\bfseries TModel 2} & 0.4998 & 0.5794 & 1.191 & 1.39 &15 & $15^\circ$--$26^\circ$  & $5^\circ$ \\

{\bfseries TModel 3} & 0.5063   &   0.5456 &     0.9906  &     1.039 & 15 &  $12^\circ$--$26^\circ$ & $5^\circ$ \\

\hline
\end{tabular}
\caption{Different models of the transition region are studied for a
  fixed black hole spin of $a/M=0.9$.  Otherwise similar to Table~\ref{tbl2}. }
\label{tbl3}
\end{center}
\end{table*}

  \subsection*{TModel 2}

  The second model, called ``TModel 2,'' uses equation (\ref{model4})
  for $\Omega_F$ with $\Omega_{\rm DBH}/\Omega_H = 0.32$, $r_{\rm
    trans\Omega}=3r_+$, and $r_0=2r_+$, and equation (\ref{transB}) with
  $r_{\rm transB}=2r_+$.  This model can be used to study any $a/M$
  and has a smooth transition from a Keplerian $\Omega_F$ at large
  radii to a fixed $\Omega_F$ near the black hole.  TModel 2 leads to
  qualitatively and quantitatively similar results as TModel 1.

  \subsection*{TModel 3}

  The final model, called ``TModel 3,'' uses equation (\ref{model3}) for
  $\Omega_F$ with $\Omega_{\rm DBH}/\Omega_H = 0.59$ and $r_{\rm
    trans\Omega}=2r_+$, and again equation (\ref{transB}) with $r_{\rm
    transB}=r_+$.  This last model is chosen to match most closely to the
  GRMHD transition regardless of whether additional discontinuities are
  created in the magnetosphere.

  Most features of this model are qualitatively similar, such as the
  field lines as shown in Figure~\ref{a.9375_t0_tf} and the power output
  distribution as shown in Figure~\ref{energyflux}. There are two new
  qualitative results, however.  First, there are additional current
  sheets in the magnetosphere near the disk inside the ergosphere. This
  is due to $\Omega_F$ not matching between the disk and the black hole.
  These additional currents indicate that no simple force-free solution
  can replace the matter-dominated disk.  The second feature is that the
  field strength at the poles of the black hole is about $2.5\times$
  more enhanced compared to the initial field strength. This field
  enhancement and the faster disk rotation leads to about $5\times$ more
  power output from the black hole and about $2\times$ more power from
  the disk if one normalizes the power by the disk field strength.  This
  suggests that an accurate model of the transition region, only
  possible in full GRMHD, is important in obtaining a quantitative
  estimate for the power output.

  \subsection*{Transition Model Summary}

  In summary, the accreting matter plays a significant role in setting
  up the transition region, and no trivial normalization of the
  force-free results can provide an unbiased estimate of the power
  output.  In GRMHD models, the disk+corona absorbs much of the power
  that reaches large distances in the GRFFE models.  Force-free models
  with a thick wedge to replace the disk+corona may better reproduce the
  power lost to the disk and corona.

  However, we found that the qualitative behavior of the force-free
  Poynting-jet from the black hole does not depend on the details of the
  matter flow in the transition region between the disk and black hole.
  For a good estimate of the power output that would correspond to the
  power output found in GRMHD numerical models, one can truncate the
  power output in the GRFFE model to include that power only above the
  disk+corona scale height in the funnel that contains the
  Poynting-dominated jet.  For the GRMHD models studied here, this angle
  away from the disk is $H/R\approx 0.6$.  To convert the power output
  to physical units, one can use the dependence of the horizon magnetic
  field strength on the mass accretion rate and black hole spin found in
  GRMHD models \citep{mckinney2005a}.

  Quantitative estimates of the Lorentz factor and opening angle are
  insensitive to the details of the transition region.  This is because
  with a $\nu=3/4$ current distribution, the poloidal field direction
  does not significantly change for any changes in the disk or black
  hole angular frequency.

  \subsection{Comparison of GRMHD and GRFFE $a/M=0.9375$ Models}\label{compare}

\begin{figure}
    \includegraphics[width=3.3in,clip]{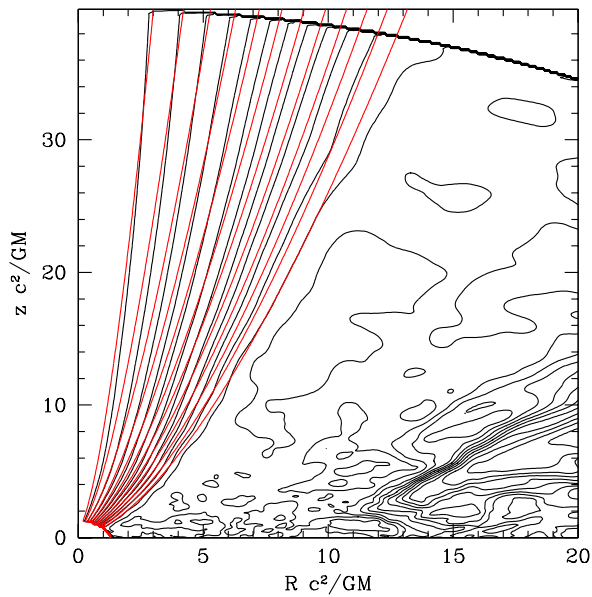}

    \caption{Overlay of the time-averaged poloidal field from the
      GRMHD numerical model of an accretion disk with a rapidly
      spinning black hole with $a/M=0.9375$ (black) and the poloidal
      field from the numerical GRFFE model with $\nu=3/4$ and the same
      value of $a/M$ (red).  Only the portion of the $\nu=3/4$ GRFFE
      solution that overlaps the funnel region of the GRMHD model is
      shown.  Note the excellent agreement between the two models in
      the overlap region.  However, the agreement is no better than
      with the non-rotating $\nu=3/4$ solution shown in Figure 9 of
      McKinney \& Narayan (2006).  Thus, while the choice $\nu=3/4$ is
      key to obtain a good match in the poloidal field structure, the
      rotation of the disk and the black hole play a lesser role. }
      \label{overlayfields9375} \end{figure}

  Figure~\ref{overlayfields9375} shows the field geometry from a
  time-dependent GRMHD numerical model as shown in Figure 9 from
  \citet{mn06}, but now overlayed with the $a/M=0.9375$ $\nu=3/4$
  GRFFE stationary solution for the poloidal field.  The funnel
  region, along which the Poynting-dominated jet in the GRMHD model
  emerges, has a poloidal field reasonably consistent with the
  $\nu=3/4$ current sheet solution.

  Hoop-stresses can be discarded as important in helping to collimate
  the jet since the poloidal field geometry is {\it decollimated} by
  the action of the black hole spin, leading to even larger
  differences between the GRMHD and GRFFE poloidal fields near the
  poles compared to the differences found in \citet{mn06} where we
  compared the polar field from the GRMHD solution and the
  non-rotating $\nu=3/4$ force-free solution.

  The field lines are more collimated in the GRMHD model than the
  GRFFE model for otherwise similar models.  This is likely due to a
  small extra pressure support by the corona or coronal wind.  Both
  polar axes show the same behavior, so this is not likely due to
  residual time-dependence in the GRMHD model.  Despite this effect,
  the opening angle for most of the power output at large distances is
  similar for the GRMHD and GRFFE models.  This means in the GRMHD
  models that the coronal wind provides some additional collimation,
  but the primary confinement of the Poynting-dominated jet is due to
  the corona.  In the GRMHD models, the gas+magnetic pressure of the
  corona that contains no ordered field is found to be similar to the
  pressure provided by the force-free field in GRFFE models.  This
  agreement between the pressures explains why the funnel field
  geometry is similar, but the process that sustains this balance
  is unknown and should be studied.


  \section{GRMHD disk model with initial $\nu=3/4$ GRFFE field}\label{GRFFE2GRMHD}

  In \citet{mn06} we showed that the $\nu=3/4$ power-law model
  represents well the currents measured in fully turbulent GRMHD
  numerical models.  We have also in this paper shown that the
  magnetic field geometry and the power output in the
  Poynting-dominated funnel region near the poles of the black hole
  are properly modelled by the $\nu=3/4$ rotating force-free model.
  However, in the GRMHD models there are no ordered field lines that
  thread the disk or corona.  What happens to these field lines?

  In this section we study the process whereby the turbulence-driven
  convective and magnetic instabilities in the disk and corona destroy
  an ordered field.  This process was discussed in
  \citet{mg04,mckinney2005a}, where it was indicated that the final
  geometry of the field lines was nearly independent of the initial
  field geometry.  The initial field geometries in those studies
  include a single field loop, many field loops with different
  poloidal directions, and a uniform net vertical field.  In all
  cases, the final field geometry within the accretion disk, corona,
  and around the black hole was qualitatively similar (see also
  \citealt{mckinney2005a}).  Here we perform a similar study with the
  non-rotating $\nu=3/4$ solution as the initial field geometry.

  \subsection{Numerical Setup}\label{setup}

  The GRMHD equations of motion are integrated numerically using a
  modified version of a numerical code called HARM \citep{gmt03}, which
  uses a conservative, shock-capturing scheme. Compared to the original
  HARM, the inversion of conserved quantities to primitive variables is
  performed by using a new faster and more robust two-dimensional
  non-linear solver \citep{noble06}.  A parabolic interpolation scheme
  \citep{colella84} is used rather than a linear interpolation scheme,
  and an optimal total variational diminishing (TVD) third order
  Runge-Kutta time stepping \citep{shu97} is used rather than the
  mid-point method.  For the problems under consideration, the parabolic
  interpolation and third order time stepping method reduce the
  truncation error significantly, even in regions where $b^2/\rho_0\gg
  1$.

  \subsection{GRMHD Model with $\nu=3/4$ Initial Field}

  The $\nu=3/4$, $a/M=0$ solution for the poloidal field is implanted in
  the same torus model of the accretion disk as studied in \citet{mg04}
  and the GRMHD equations are integrated.  As in the fiducial model of
  \citet{mg04}, the black hole has a spin of $a/M=0.9375$, which is
  close to the equilibrium value of $a/M\sim 0.92$ \citep{gsm04}.

  \begin{figure}
    \includegraphics[width=3.3in,clip]{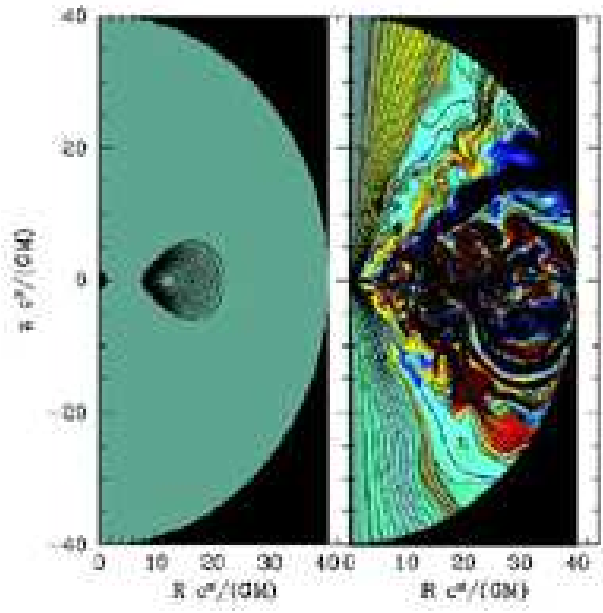}

    \caption{ Initial (left panel) and final (right panel) poloidal
      magnetic field geometry (solid black lines) and toroidal field
      strength ($B_\phi$, red positive, blue negative) for a GRMHD
      numerical model that was initiated with a gas torus containing a
      poloidal magnetic loop. The model has a rapidly spinning black
      hole with $a/M=0.9375$. The panel on the right shows the standard result for a GRMHD
      numerical model of an accretion disk, in which the final system
      has a disk, corona and outflowing wind with disordered field,
      and a polar region (the Poynting-dominated jet) with highly
      ordered field. } \label{grmhd_ic_fc} \end{figure}

  As a reference point, Figure~\ref{grmhd_ic_fc} shows the initial and
  final field geometry for the GRMHD model with $a/M=0.9375$ with an
  initial field corresponding to a single field loop in the torus, as
  studied by \citet{mg04} and others.

  \begin{figure}
    \includegraphics[width=3.3in,clip]{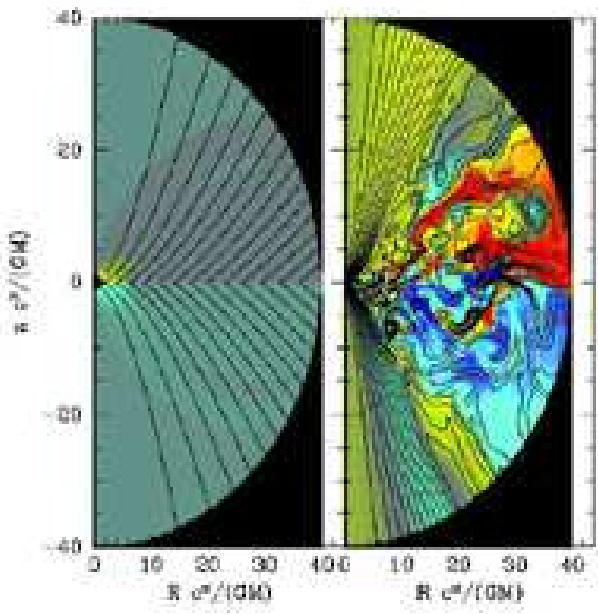}

    \caption{Similar to Figure~\ref{grmhd_ic_fc} except that the initial magnetic field
      corresponds to the $\nu=3/4$ force-free solution.  Despite the
      very different initial conditions, the final states in the two
      models are similar. }
      \label{grmhd_ic_fc_nfourth}
  \end{figure}

  Figure~\ref{grmhd_ic_fc_nfourth} shows the initial and final field
  geometry for the GRMHD model with $a/M=0.9375$ that uses the
  $\nu=3/4$ force-free solution for the initial field geometry.  A
  model initially with a net vertical field leads to qualitatively
  similar results.  The GRMHD model with a single loop exhibits more
  turbulence and larger variations in the polar enclosed poloidal
  current, which at high/low levels saturates to red/black in the
  color figure.

  The final poloidal field distribution at the poles follows the
  $\nu=3/4$ force-free solution for both the models as shown in
  Figures~\ref{grmhd_ic_fc} and~\ref{grmhd_ic_fc_nfourth}. Other
  models, not shown, such as one with an initial net vertical field
  also agree with the $\nu=3/4$ force-free solution in the polar
  region.  The value of $\Omega_F$ in the disk and transition region
  is also similar.  This is despite the fact that the detailed
  structure of the magnetic geometry in the disk is different.  For
  every model we have studied, the turbulence due to the
  magnetorotational instability efficiently redistributes the currents
  to the $\nu=3/4$ power-law distribution.

  For these models, the time evolution shows that the global magnetic
  field bifurcates into two regions.  The funnel region forms as
  material with low angular momentum falls into the black hole and
  launches a jet, while many of the ordered fields threading the torus
  are tied to material that is gaining angular momentum and so is
  launched backwards along the disk surface.

  As the bifurcation unfolds, the pressure support of the $\nu=3/4$
  field is lost in the region outside the funnel.  As the disk accretes,
  some of the disk surface material resupplies that evacuated region to
  replace the pressure that was provided by the field.  This process
  forms the corona and drives the disk wind.

  A measurement of the gas+magnetic pressure in the coronal region shows
  that the evolved GRMHD model with the corona has a similar pressure as
  the initial force-free fields in that region.  There is a small
  additional pressure driven by the flux of mass, and this explains why
  the GRMHD model is slightly more collimated than the force-free
  $\nu=3/4$ model.

  A similar quasi-stationary state of the accretion disk and field is
  found with various other initial field geometries
  \citep{hirose04,mckinney2005a}.  Only a contrived geometry with a
  purely toroidal field leads to significant deviations in the
  quasi-stationary state, and of course models with a purely toroidal
  field have no toroidal currents.

  \section{Limitations}\label{limitations}

  The primary limitation of the present study is that the numerical
  models are axisymmetric.  Therefore, while the GRFFE simulations show
  that the Poynting-dominated jet is stable to axisymmetric pinch
  perturbations, we cannot tell whether it is kink stable.

  Another limitation is that in the GRFFE models the disk is assumed to
  be infinitely thin.  This is in contrast to the GRMHD models where the
  disk is thick resulting in some power from the black hole being
  absorbed by the disk and corona. However, because the force-free
  region in the GRMHD simulations is sharply defined, a good estimate of
  the power can be obtained from the GRFFE models by limiting the
  integration of the angular power density to a cone around the poles
  that is the same angular size as that found in the GRMHD models.  In
  this way, the Poynting-dominated jet power can be estimated for any
  disk thickness.

  \section{Conclusions}\label{conclusions}

  Motivated by the simple toroidal current distribution we found in
  GRMHD numerical models of turbulent accretion disks \citep{mn06}, in
  this paper we have studied simple force-free (GRFFE) models of the
  black hole and disk magnetosphere.  We idealized the GRMHD solution
  for the accretion disk as an equatorial rotating conductor with
  boundary conditions defined by an equatorial toroidal current and
  field angular frequency.

  To begin with, in \S~3 we considered the paraboloidal problem of
  \citet{bz77} in which the current is of the form $dI_\phi/dr \propto
  r^{-1}$.  For a slowly rotating black hole and a non-rotating disk,
  we obtained excellent agreement between the numerical GRFFE solution
  and BZ's analytical solution; this is not surprising since the BZ
  solution was derived in this limit.  We then considered a rapidly
  rotating black hole with a non-rotating disk and also a rotating
  hole with a Keplerian disk.  In these cases, the GRFFE model
  deviates to some extent from the BZ solution, but the agreement
  between the two is still reasonably good.  For all models,
  collimation was found to be primarily due to the poloidal field
  geometry determined by the toroidal current in the disk.  Forces due
  to hoop-stresses nearly cancel centrifugal forces.

  In paper I, we considered a self-similar force-free model of an
  equatorial current sheet with a toroidal current following
  $dI_\phi/dr \propto r^{-5/4}$ (the $\nu=3/4$ model) \citep{mn06}. In
  that paper we showed the structure of the field from the force-free
  solution agrees well with the poloidal field in the jet region of
  GRMHD numerical models.  However, because the force-free model had
  no rotation, it had no toroidal field and we were unable to compare
  the toroidal structure or the Lorentz factor of GRMHD jets.

  In \S~4, we considered a similar force-free model with $dI_\phi/dr
  \propto r^{-5/4}$ in the equatorial plane, but with an arbitrarily
  rapidly rotating black hole and a Keplerian disk.  One well-known
  subtlety of force-free models of the black hole and disk is how to
  define quantities within the transition region between the black
  hole and accretion disk \citep{bz77}.  We experimented with two
  different models, motivated by GRMHD simulations, for the field
  angular velocity in the transition region.

  The agreement between these improved $\nu=3/4$ GRFFE models and the
  jet/funnel region of the GRMHD models is excellent.  We found that
  the poloidal structure of the field in the rotating GRFFE model is
  nearly the same as in the non-rotating model considered in
  \citet{mn06}, and like the latter agrees well with the poloidal
  structure of the field in the GRMHD jet.  Within the jet driven by
  the black hole, forces due to hoop-stresses nearly cancel
  centrifugal forces, suggesting that self-collimation is not
  important. The toroidal field structure in the rotating GRFFE model
  agrees very well with the toroidal structure of the GRMHD jet, as do
  the power output in the jet and the Lorentz factor of the jet. Thus,
  the rotating GRFFE model gives an accurate description of the
  acceleration, collimation, jet opening angle, field pitch angle,
  electromagnetic power, etc., of the relativistic jet seen in GRMHD
  simulations.

  While there is excellent agreement in the jet regions of the two
  solutions, there are large differences between the force-free and
  GRMHD models in other regions.  In the force-free model the primary
  agent responsible for collimating the jet is the poloidal field
  threading the disk.  These field lines lie external to the field
  lines in the jet (which thread the black hole), and it is the
  pressure of this external field that produces the collimation. In
  the GRMHD numerical model, on the other hand, the region external to
  the jet is filled with a disorganized, magnetized corona and wind,
  and it is the pressure from this gas that causes the collimation.
  This suggests that the corona is required for the collimation of the
  Poynting-dominated jet in GRMHD simulations.

  We also found that the GRFFE jet is stable to axisymmetric Eulerian
  perturbations. In the GRFFE model, the black hole jet is in
  force-balance with the force-free disk wind.  In the GRMHD models of
  \citet{mckinney2006c}, we found that at large radii the jet is {\it
  unstable} to axisymmetric perturbations. This is consistent with the
  lack of coronal material (and so coronal pressure support that
  replaces the magnetized support in the force-free regime) at large
  radii.  This suggests that astrophysical jets require support by a
  corona or disk wind in order to remain stable, but at large radii
  such support may be naturally unavailable and this may lead to
  toroidal-field-driven instabilities and dissipation in the jet as
  shown in \citet{mckinney2006c}.

  The lack of an ordered field in the corona is related to the large
  difference between the GRFFE and GRMHD models with respect to the
  power output from the disk.  In the force-free model, the disk field
  lines carry out electromagnetic power in the form of Poynting flux.
  In the GRMHD model, however, little electromagnetic power from the
  disk reaches large radii.  Instead, most of the electromagnetic
  energy from the disk is dissipated in the matter and converted to
  other forms of energy in the plasma \citep{mn06}.

  In \S~6, we described a numerical experiment in which we further
  explored the connection between the $\nu=3/4$ ordered GRFFE solution
  and the magneto-rotationally unstable turbulent GRMHD model.  An
  accretion disk was embedded with the $\nu=3/4$ ordered field and
  followed using the GRMHD code. During the evolution of the system,
  the toroidal current in the disk retains its initial $\nu=3/4$
  behavior.  However, the large-scale fields associated with the
  initial state are driven outwards by angular momentum transport and
  the pressure of those fields is replaced by coronal gas+magnetic
  pressure.  This suggests that there is a continuous coupling between
  the currents in the disk and the coronal material.  The corona and
  coronal wind are generated by reconnection processes and magnetic
  stresses in the disk.  Even without an organized field the coronal
  wind is collimated, as suggested by \citet{hb00,li2002}. The
  gas+magnetic pressure in the corona is similar to the magnetic field
  pressure in the GRFFE model, and it is because of this that the jet
  regions are so similar in the two models.  It remains unknown by
  what mechanism the pressure of the corona in GRMHD models drives the
  power-law $\nu=3/4$ toroidal current that generates the
  corresponding magnetized jet.

  \section*{Acknowledgments}

  This research was supported by NASA-Astrophysics Theory Program
  grant NNG04GL38G and a Harvard CfA Institute for Theory and
  Computation fellowship.  I thank Dmitri Uzdensky, Serguei
  Komissarov, and Vasily Beskin for useful comments and discussions.


  \appendix

  \section{Asymptotic Form of Pitch Angle and Lorentz Factor}\label{lorentzscaling}

  In this appendix, we show how the Lorentz factor scales with radius
  for a given power-law distribution of currents in the disk.  First, we
  remind the reader that the general relativistic language can be
  translated into an special relativistic language (3+1 splitting) by
  defining a space-time foliation.  A convenient foliation is to
  choose to measure all quantities with respect to an observer with
  4-velocity of a zero angular momentum observer (ZAMO) frame.  This
  is given by $\eta_\mu=\{-\alpha,0,0,0\}$, where
  $\alpha=1/\sqrt{-g^{tt}}$.  In this frame, the force-free equations
  take the same form as in Minkowski space-time.

  In the ZAMO frame the drift velocity is written as
  \begin{equation}
    v^i = \frac{\epsilon^{ijk} E_j B_k}{B^2} ,
  \end{equation}
  where $\epsilon^{ijk}$ is the spatial permutation tensor.  The
  Lorentz factor is given by
  \begin{equation}
    \Gamma = \frac{1}{\sqrt{1-v^2}} ,
  \end{equation}
  as in special relativity. The electric field is given by
  \begin{equation}\label{EFIELD}
    E^i = -\epsilon^{ijk} v_j B_k ,
  \end{equation}
  Given flux-freezing, axisymmetry, and stationarity, the fluid velocity
  or drift velocity is related to the magnetic field and field angular
  frequency via the relation
  \begin{equation}\label{fluxfreeze}
    \frac{v_p}{B_p} = \frac{v_t - R \Omega_F}{B_t}
  \end{equation}
  where these are orthonormal basis quantities where $p$ stands for
  poloidal, $t$ stands for toroidal, and $R$ is cylindrical radius
  \citep{bek78}. For a review of the force-free equations see
  \citet{mckinney2006a}.

  Now, consider the asymptotic regime where $r\rightarrow \infty$, and
  assume that the toroidal field dominates in this regime such that
  the mostly radial poloidal velocity approaches the speed of light.
  Then from equation (\ref{EFIELD}) one finds that
  \begin{equation}
    |E| \approx |E^\theta| \approx |B_t| .
  \end{equation}
  From the drift velocity, one has that
  \begin{equation}
    v^2 = \frac{|\mathbf{E}\times\mathbf{B}|^2}{B^2 (B_p^2+B_t^2)^2}
    \approx \frac{E^2}{(B_p^2+B_t^2)^2} \approx \frac{1}{(B_p/B_t)^2+1} ,
  \end{equation}
  and so the Lorentz factor can be written as
  \begin{equation}
    \Gamma \approx \frac{B_t}{B_p} .
  \end{equation}
  Beyond the light cylinder, $v_t\ll R\Omega_F$ and assuming $v_p$
  approaches the speed of light, equation (\ref{fluxfreeze}) then gives
  that
  \begin{equation}\label{equationbtbp}
    \frac{B_t}{B_p} = -R \Omega_F .
  \end{equation}
  Notice that this result is consistent with the numerical results of
  this paper, where we discuss the pitch angle such as given by
  equation (\ref{pitchangle}).

  The above implies that
  \begin{equation}
    \Gamma = R\Omega_F  = \Omega_F r^{1-\alpha} ,
  \end{equation}
  for a field geometry that follows $\theta_j \propto r^{-\alpha}$.
  In general, for a solution with a current distribution with
  power-law index $\nu$, in the slow-rotation approximation, the
  minimal-torque solution \citep{michel69} gives an opening angle of
  approximately
  \begin{equation}
  \theta_j\propto r^{-\nu/2} ,
  \end{equation}
  and so
  \begin{equation}
  \Gamma\propto r^{1-\nu/2} .
  \end{equation}
  For example, for the paraboloidal field geometry with $\nu=1$, one
  has that $\theta_j\propto r^{-1/2}$, so that $\Gamma\propto
  r^{1/2}$.  For $\nu=3/4$ one has that $\theta_j\propto r^{-0.375}$
  so that $\Gamma\propto r^{0.625}$.  This basic scaling is fairly
  consistent with the results of this paper, where deviations are
  expected for a general black hole spin and disk rotation profile.
  See also \citet{beskin97,beskin98,bm00,bn06,nmf06}.



  \label{lastpage}

\end{document}